\documentclass[%
 reprint,
nofootinbib,
 amsmath,amssymb,
 aps,
]{revtex4-2}
\usepackage[dvipsnames,table,xcdraw]{xcolor}
\definecolor{hworange}{HTML}{DD1C77}
\definecolor{nrblue}{HTML}{3182BD}
\usepackage{soul}
\usepackage{graphicx}
\usepackage{dcolumn}
\usepackage{bm}
\usepackage{caption}
\usepackage{hyperref}
\usepackage{booktabs}
\hypersetup{
    colorlinks = true,
    linkcolor = {hworange},
    citecolor = {nrblue},
    urlcolor = {magenta},
}

\usepackage{braket}

\usepackage[linesnumbered,ruled,vlined]{algorithm2e}
 \usepackage{orcidlink}

\usepackage{tabularx}
\usepackage{amsmath}
\usepackage{bm}
\usepackage{hyperref}
\usepackage{float}
\usepackage{rotating}
\usepackage{subcaption}
\usepackage[normalem]{ulem}
\usepackage[justification=raggedright]{caption}
\usepackage{multirow}

\usepackage{tikz}
\usetikzlibrary{quantikz2}

\usepackage{enumitem}

\usepackage[inkscapelatex=false]{svg}

\newcommand{\R}[1]{\textcolor{red}{ref}}

\begin{document}

\preprint{APS/123-QED}

\title{Quantum Generative Adversarial Autoencoders:\\ Learning latent representations for quantum data generation}%

\author{Naipunnya Raj \orcidlink{0009-0009-1692-114X}}
\thanks{These two authors contributed equally. \\
Emails: \href{mailto:naipunnya.raj@fujitsu.com}{naipunnya.raj@fujitsu.com} ; \href{mailto:rajiv.sangle@fujitsu.com}{rajiv.sangle@fujitsu.com}}

\author{Rajiv Sangle \orcidlink{0009-0007-7716-6315}}
\thanks{These two authors contributed equally. \\
Emails: \href{mailto:naipunnya.raj@fujitsu.com}{naipunnya.raj@fujitsu.com} ; \href{mailto:rajiv.sangle@fujitsu.com}{rajiv.sangle@fujitsu.com}}

\author{Avinash Singh \orcidlink{0009-0009-9116-2884}}

\author{Krishna Kumar Sabapathy \orcidlink{0000-0003-3107-6844}}
\affiliation{%
\normalsize Quantum Lab, Fujitsu Research of India
}

\begin{abstract}

In this work, we introduce the Quantum Generative Adversarial Autoencoder (QGAA), a quantum model for generation of quantum data. The QGAA consists of two components: (a) Quantum Autoencoder (QAE) to compress quantum states, and (b)
Quantum Generative Adversarial Network (QGAN) to learn the latent space of the trained QAE. This approach imparts the QAE with generative capabilities. The utility of QGAA is demonstrated in two representative scenarios: (a) generation of pure entangled states, and (b) generation of parameterized molecular ground states for $\text{H}_{2}$ and $\text{LiH}$. The average errors in the energies estimated by the trained QGAA
are
$0.02$ Ha for $\mathrm{H}_2$ and $0.06$ Ha for $\mathrm{LiH}$ in simulations upto 6 qubits. These results illustrate the potential of QGAA for quantum state generation, quantum chemistry, and  near-term quantum machine learning applications. 

\end{abstract}

\maketitle


\section{Introduction} 
Over the past decade, 
machine learning has undergone transformative advancements, primarily fueled by the development of sophisticated deep learning architectures and training methodologies. In parallel, Quantum Machine Learning (QML) has emerged as a field dedicated to exploring how quantum algorithms and quantum computing platforms can be utilized to process, model, and extract meaningful insights from data \,\cite{biamonte2017quantum, cerezo2021variational, wang_2024}, and also generate new data \cite{sengar2025generative, gan_original}. While efforts in QML primarily focused on leveraging quantum computing to accelerate classical machine learning tasks \cite{vision_trasformer,Khoshaman_2019}, a significant and increasingly important direction involves the development of quantum models that operate directly on quantum data \cite{biamonte2017quantum, qgan_01, bhat2025meta}. These models, tailored specifically to quantum data, are essential for realizing the full potential of quantum technologies, enabling applications in quantum information processing that are intractable with classical methods \cite{DEVADAS2025103318}.

A notable model within QML for handling quantum data is the Quantum Autoencoder (QAE), which draws inspiration from its classical counterpart, the Autoencoder (AE) \,\cite{ae_hinton_learning_internal_representations, ae_review}. QAE has been applied to demonstrate how quantum circuits can be trained to compress quantum states, with applications to quantum simulation and quantum information \,\cite{romero2017quantum, huang2020realization, qae_compression_bounds, cao2021noise, ma2024quantum}. Further developments extend these architectures to the denoising of entangled quantum states under realistic noise models \,\cite{PhysRevLett.124.130502, PhysRevResearch.6.023181,achache2020denoising, ae_denoising},  along with proposals for error mitigation strategies tailored to Noisy Intermediate-Scale Quantum (NISQ) devices \,\cite{zhang2021generic, mok2024rigorous}. Practical realizations of QAE in quantum hardware, such as nitrogen-vacancy centers, demonstrated robust compression and the preservation of entanglement, while significantly lengthening the coherence times of Bell states \,\cite{entanglement_AE}. 

Despite these promising applications, it is important to note that the QAE, on its own, does not possess generative capabilities. It is fundamentally designed to learn a low-dimensional latent space that captures the essential features of a given dataset \cite{romero2017quantum}. However, this compression process is not inherently generative since it does not allow direct access to the quantum latent space.
In comparison, the classical Variational Autoencoder (VAE) enables direct access to the latent space of an Autoencoder by regularizing the latent space with a known distribution. This allows sampling from the latent space to generate new data samples \cite{bacarreza2025quantum}.
There is no straightforward natural analogue to access the quantum latent space of a QAE for quantum generative tasks.

Alongside these developments, the intersection of classical generative models and QML has catalyzed significant interest in the formulation and development of quantum generative models, designed to learn and generate quantum data using quantum processes\,\cite{tian2023recent, vishnu2025density}. Notable examples include the Quantum Generative Adversarial Network (QGAN) \,\cite{qgan_01, qgan_02, eqgan, hamiltonian_qgan}, Quantum Boltzmann Machine \,\cite{amin2018quantum}, Quantum Circuit Born Machine \,\cite{coyle2020born}, Quantum Transformers \cite{kamata}  and Quantum Generative Diffusion Model \,\cite{quantum_diffusion_model}.
These quantum generative models are formulated to learn and synthesize quantum states.
In particular among these, works on the QGAN \cite{qgan_01, qgan_02, eqgan, hamiltonian_qgan} establish theoretical foundations for quantum adversarial learning for quantum generative tasks, inspired by the formalism of 
classical Generative Adversarial Networks (GAN) \cite{gan_original}. 

To address the lack of generative capability of QAE, we propose a novel QML architecture, the Quantum Generative Adversarial Autoencoders (QGAA). It combines the compression power of the QAE with the generative capabilities of the QGAN. We demonstrate how the adversarial training framework of QGAN can be used to learn the latent quantum representation encoded by a trained QAE, imparting the model with generative capabilities. The key contributions of this work are as follows:

\begin{enumerate}
    
    \item We present the theoretical framework of quantum adversarial learning to learn the representation of the quantum latent space of a trained QAE. 
    The adversarial learning approach gives the ability to directly generate new quantum latent states.
    
    \item We demonstrate two applications:
    \begin{enumerate}
        \item Learning the latent representation of a set of 2-qubit entangled states.
    
        \item Learning the ground state energy profiles of the 
        parameterized molecular Hamiltonians
        of
        $\mathrm{H}_2$ (4 qubits) and $\mathrm{LiH}$ (6 qubits).

    \end{enumerate}
    These tasks allow us to evaluate the performance of QGAA while also highlighting practical challenges encountered during implementation.

\end{enumerate}

A key advantage of the QGAA approach lies in its potential to reduce quantum  resource requirements 
by compressing quantum states into a lower-dimensional latent space, while simultaneously enabling enhanced generative capabilities through adversarial learning.

The paper is organized as follows: Section \ref{sec:background} introduces the QAE, which finds the compressed representation of quantum states, and QGAN, a quantum generative model.
Section \ref{sec:adversarial_formalism} details the quantum adversarial framework for learning the latent space representation of a trained QAE. Sections \ref{sec:example_1} and \ref{sec:example_2} demonstrate the implementation of this architecture 
for two different quantum-native generative tasks.
Section \ref{sec:conclusion} concludes the article with insights gained from this work and suggests future research directions.
\section{Background}

\label{sec:background}

\begin{figure*}[!htb]
\centering
    \includegraphics[width=\linewidth]{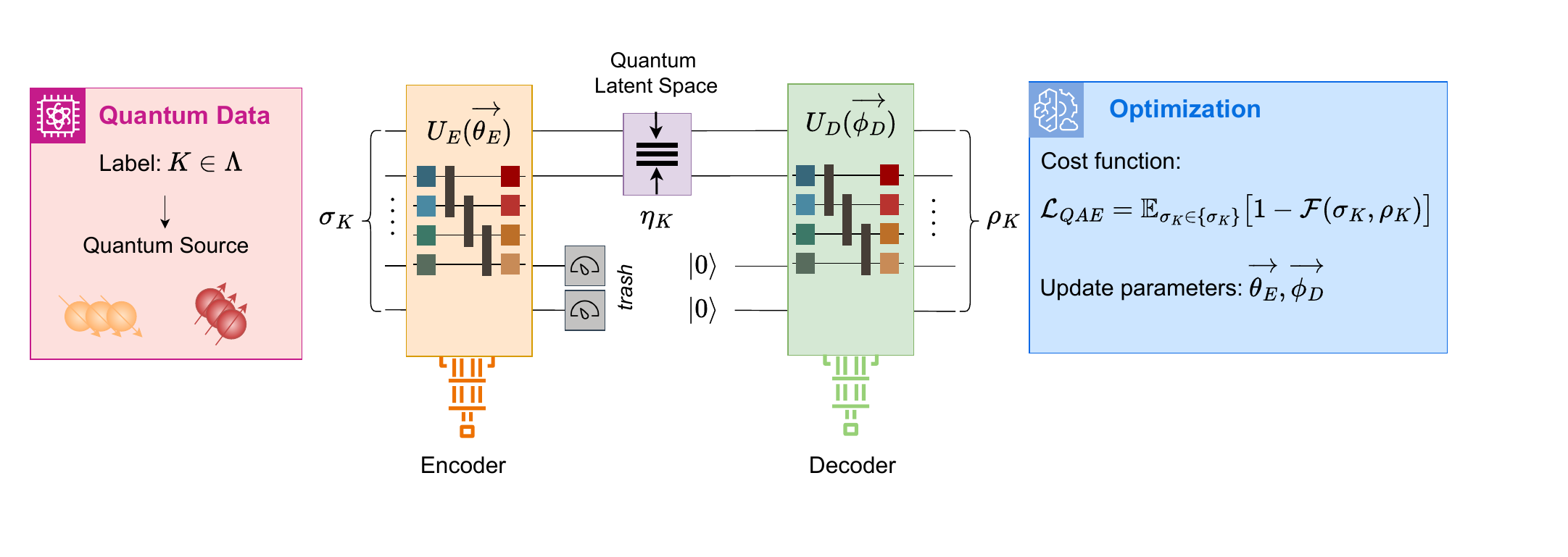}
    \caption{\textbf{Compression stage using QAE:} The input quantum states \( \sigma_{K} \) is indexed by some label $K \in \Lambda$ which uniquely characterizes the state $\sigma_{K}$.  Here, $\Lambda := \{ K \}$ denotes the set of all valid labels that define $\{ \sigma_{K} \}$. The encoder applies a parametrized unitary \( U_E(\overrightarrow{\theta_E}) \) to each input state \( \sigma_{K} \), generating an entangled intermediate state. A designated subset of qubits referred to as the \textit{trash}, is then traced out to yield the compressed latent state \( \eta_K \). The decoder subsequently applies a second parametrized unitary \( U_D(\overrightarrow{\phi_D}) \) to reconstruct the output state \( \rho_{K} \). The parameters \( \overrightarrow{\theta_E} \) and \( \overrightarrow{\phi_D} \) are trained to minimize the reconstruction loss, $\mathcal{L}_{QAE}$.}
    \label{fig:compression}
\end{figure*}

In this section, we provide an overview of the two QML models that form the foundation of this work: (i) the Quantum Autoencoder for the compression stage and (ii) the Quantum Generative Adversarial Network for the generative training stage.

\subsection{Quantum Autoencoder}
\label{sec:background_qae}

Autoencoder (AE) is a classical machine learning model designed to compress data into a low-dimensional latent space and reconstruct the compressed data to its original structure \cite{kingma2013auto}. Building on the principles of the AE, the Quantum Autoencoder (QAE) was introduced as a quantum framework for efficient compression and reconstruction of quantum states \cite{romero2017quantum}. An overview of classical AE is provided in Appendix \ref{sec:ae_appendix}. This section details the architecture and the training process, designed to minimize information loss during compression and reconstruction of quantum states by QAE, as illustrated in Figure \ref{fig:compression}.

The core concept of a QAE is learning an optimal compressed or latent form of a given set of input quantum states $\Gamma := \{ \sigma_{K} \}$. 
Each state, $\sigma_{K} \in \Gamma$, is uniquely characterized
by some label, $K \in \Lambda$.
Here, $\Lambda := \{ K \}$ denotes the set of all valid labels that define $\Gamma$.

\textbf{Architecture:} The QAE consists of three key components. The encoder is a parametrized unitary transformation
\begin{equation}
    U_E(\overrightarrow{\theta_E}): \mathcal{H}_A \longrightarrow \mathcal{H}_L \otimes \mathcal{H}_T,
\end{equation}
acting on input states \(\sigma_{K} \in \mathcal{D}(\mathcal{H}_A)\), 
where \(\mathcal{D}(\mathcal{H})\) denotes the set of density operators on the Hilbert space \(\mathcal{H}\). 
Here, \(\overrightarrow{\theta_E}\) represents the trainable parameters, 
\(\mathcal{H}_A \cong (\mathbb{C}^2)^{\otimes n}\) is the $n$-qubit Hilbert space of the input, 
\(\mathcal{H}_L \cong (\mathbb{C}^2)^{\otimes \ell}\) is the latent subspace of dimension \(2^\ell\), and 
\(\mathcal{H}_T \cong (\mathbb{C}^2)^{\otimes (n-\ell)}\) is the trash subspace of dimension \(2^{\,n-\ell}\). 
Tracing out the trash subsystem yields the compressed latent state

\begin{equation}
    \eta_{K} = \mathrm{Tr}_T \big[ U_E(\overrightarrow{\theta_E}) \, 
    \sigma_{K} \, 
    U_E^\dagger(\overrightarrow{\theta_E}) \big] 
    \in \mathcal{D}(\mathcal{H}_L).
\end{equation}

The decoder is a second parametrized unitary
\begin{equation}
    U_D(\overrightarrow{\phi_D}): \mathcal{H}_L \otimes \mathcal{H}_T \longrightarrow \mathcal{H}_A,
\end{equation}
governed by trainable parameters \(\overrightarrow{\phi_D}\). 
It acts on the latent state \(\eta_{K}\), together with an initialized trash register, 
to produce the reconstructed output state 
\(\rho_{K} \in \mathcal{D}(\mathcal{H}_A)\). The unitaries \(U_E(\overrightarrow{\theta_{E}})\) and \(U_D(\overrightarrow{\phi_{D}})\) are typically realized as parameterized quantum circuits, composed of single-qubit rotation gates (with trainable angles) and fixed entangling gates.

\textbf{Training:} 
The objective is to minimize the loss function that quantifies the distance between the original and reconstructed quantum states in a given data ensemble.

In practice, the SWAP test \cite{swap_00, swap_quantum_fingerprinting, swap_switch_test} is employed to estimate the overlap between two quantum states. If \(\sigma_{K}=\ket{\psi_{K}}\bra{\psi_{K}}\) and $\rho_{K} = \ket{\phi_{K}}\bra{\phi_{K}}$ are pure states, then the fidelity of the two states is equal to the overlap evaluated using the SWAP test, i.e., $\mathcal{F}(\sigma_{K}, \rho_{K}) = \text{SWAP} (\sigma_{K}, \rho_{K}) = |\braket{\psi_{K}|\phi_{K}}|^2$. In this case, the results of the SWAP test between \( \{ \sigma_{K} \} \) and \( \{ \rho_{K} \} \) are used to evaluate the QAE loss function \( \mathcal{L}_{\text{QAE}} \). 

If the standard SWAP test is applied to two mixed states, the outcome corresponds to the overlap $\mathrm{Tr}(\rho\sigma)$ rather than the fidelity. In other words, the SWAP test provides a direct estimate of fidelity only for pure states. For mixed states, the fidelity must be evaluated analytically. A detailed definition of the SWAP operator, the associated protocol, and the analytical expressions for fidelity in both pure and mixed state settings are provided in Appendix~\ref{metrics}.

For the purpose of the demonstrative examples discussed in this work, we have used the analytical expression of fidelity to define the cost function of the QAE,

\begin{equation}
    \label{eq:qae_cost_function}
    \; \mathcal{L}_{\text{QAE}} = \mathbb{E}_{\sigma_{K} \in \{ \sigma_{K} \}} \big[ 1 - \mathcal{F}(\sigma_{K}, \rho_{K}) \big].
\end{equation}

The loss function, \( \mathcal{L}_{\text{QAE}} \), is minimized using classical optimization techniques to find parameter values of the encoder, \( U(\overrightarrow{\theta_E}) \), and decoder, \( U_D(\overrightarrow{\phi_D}) \), such that the reconstructed state, \( \rho_{K} \), closely matches its corresponding input state, \( \sigma_{K} \). The optimal parameters, \( \overrightarrow{\theta^{*}_{E}} \) and \( \overrightarrow{\phi^*_D} \) obtained after training, are used to evaluate the performance of the QAE on unseen test data by computing
metrics such as fidelity. The classical Variational Autoencoder (VAE) extends the standard AE by enabling sampling from a probabilistic latent space, thereby combining compression with generative modeling, as detailed in Appendix~\ref{vae_appendix}. In contrast, the QAE lacks generative capabilities and does not allow sampling of new quantum states from its learned latent space. This absence of a fully realized quantum analogue motivates the integration of a QGAN to address the generative limitation.

\subsection{Quantum Generative Adversarial Network}
\label{sec:QGAN}
The notion of Quantum Generative Adversarial Learning \cite{qgan_01} was formalized taking inspiration from the classical GAN \cite{gan_original}. 
An overview of the classical GAN is provided in Appendix \ref{sec:gan_appendix}.

Quantum mechanics is inherently probabilistic  
and therefore the notion of generation of quantum data is different from that of classical data. Given access to a quantum source, $R$, described by some density matrix, $\sigma$, the quantum generative task is defined as training a quantum agent - the generator, $G$, to generate $\rho$, whose statistics are equal to that of $\sigma$, i.e., $\rho = \sigma$ \cite{qgan_01}. Here, statistics refers to the measurement outcomes of $\sigma$ and $\rho$ corresponding to any tomographically complete set of observables.

\textbf{Architecture:}
In a QGAN, the training of $G$ is enabled by another quantum agent - the discriminator, $D$, whose objective is to evaluate whether the input quantum state supplied to it is $\sigma$ (\textit{real data}) from $R$ or $\rho$ (\textit{fake data}) generated by $G$. Both $G$ and $D$ can be implemented as quantum circuits parameterized by $\overrightarrow{\theta_{g}}$ and $\overrightarrow{\theta_{d}}$ respectively. 

$D$ consists of an input register, where either $\sigma$ (\textit{real data}) or $\rho$ (\textit{fake data}) can be loaded.
Apart from the input data register, $D$ also consists of 
a probe qubit in some reference state such as $\ket{0}$.
The evolution dynamics of $D(\overrightarrow{\theta_{d}})$ cause the input state, $\sigma$ or $\rho$, to interact with the reference probe.
After this joint evolution of the input state and the reference probe, measuring the probe qubit in the Pauli-$Z$ basis, $\{ \ket{0}\bra{0}, \ket{1}\bra{1} \}$, is equivalent to the action of a binary Positive Operator-Valued Measure (POVM) $\{ \hat{T} \equiv \hat{T}
( \overrightarrow{\theta_{d}} ), \; \hat{F} = \hat{F} ( \overrightarrow{\theta_{d}} ) \}$ on the input state corresponding to the decisions of the input being \textit{real} ($+1$) or \textit{fake} ($-1$) \cite{qgan_01, qgan_02, helstrom1969quantum, wilde2013quantum}. Naturally, for any $\overrightarrow{\theta_{d}}$, $\hat{T}$ and $\hat{F}$ are positive semi-definite operators satisfying the completeness condition $\hat{T} + \hat{F} = \mathbb{I}$. 

$D$ implements its objective of distinguishing between the states $\sigma$ (\textit{real data}) and $\rho$ (\textit{fake data}) by maximizing the probability cost function $\mathcal{L}_{\text{QGAN}}$ \cite{qgan_01, qgan_02},
\begin{subequations}
\label{eq:qgan_cost_a_and_b}

\begin{equation}
    \label{eq:qgan_gen_state}
        \rho \equiv G (\overrightarrow{\theta_{g}})
        \text{ and }
        \hat{T} \equiv \hat{T}(\overrightarrow{\theta_{d}}),
\end{equation}

\begin{equation}
    \label{eq:qgan_cost}
        \underset{\overrightarrow{\theta_{g}}}{min} \;
        \underset{\overrightarrow{\theta_{d}}}{max} \;
        \mathcal{L}_{\text{QGAN}} := \frac{1}{2} \bigg[ 1 + \bigg\{ \text{Tr} \big( \hat{T} \sigma \big) - \text{Tr} \big( \hat{T} \rho \big)  \bigg\} \bigg].
\end{equation}
\end{subequations}
$\mathcal{L}_{\text{QGAN}}$ is designed in accordance with the quantum state discrimination protocol~\cite{helstrom1969quantum, wilde2013quantum} designed to maximize the probability of distinguishing between the quantum states $\sigma$ and $\rho$.
The upper bound on $\mathcal{L}_{\text{QGAN}}$ is given
by the theory of \text{Helstrom measurement} or the \text{minimum-error distinguishing measurement} \cite{helstrom1969quantum, wilde2013quantum}. 

\textbf{Training}:
The quantum generative adversarial learning game is set up as follows:
\begin{enumerate}
    \item The parameters, $\overrightarrow{\theta_{g}}$ and $\overrightarrow{\theta_{d}}$, are initialized, and $G$ generates an initial state, $\rho_{0}$. The corresponding $\mathcal{L}_{\text{QGAN}}$ is evaluated with the initial value of the operator $\hat{T} \equiv \hat{T}(\overrightarrow{\theta_{d}})$.

    \item The strategy of $D$ is to optimize $\overrightarrow{\theta_{d}}$ to produce a corresponding $\hat{T}^{\prime}$ that maximizes $\mathcal{L}_{\text{QGAN}}$ keeping the parameters of $G$ fixed.

    \item The strategy of $G$ is to optimize $\overrightarrow{\theta_{g}}$ to generate $\rho^{\prime}$ such that Tr$\{ \hat{T}^{\prime} \rho^{\prime} \}$ gets closer to Tr$\{ \hat{T}^{\prime} \sigma \}$ and minimizes $\mathcal{L}_{\text{QGAN}}$ keeping the parameters of $D$ fixed.

    \item Subsequent iterations of Step $2$ and Step $3$ correspond to the max-min optimization of $\mathcal{L}_{\text{QGAN}}$.
    Both the space of density matrices, $\{ \rho(\overrightarrow{\theta_{g}}) \}$, and the space of positive semi-definite operators, $\{ \hat{T}(\overrightarrow{\theta_{d}}) \}$, are convex and compact.
    Therefore, optimization over these spaces is possible and a  \text{unique Nash equilibrium} exists exactly at $\rho(\overrightarrow{\theta_{g}^{*}}) = \sigma$ for some optimal parameters $\overrightarrow{\theta_{g}^{*}}$ \cite{qgan_01}. 

    \item Upon convergence to the unique fixed-point of the game, $D$ \textit{either} distinguishes between $\sigma$ and $\rho(\overrightarrow{\theta_{g}^{*}})$ as good as a fair-coin toss \cite{qgan_01} \textit{or} identically concludes that the state is \textit{real} no matter whether the input state is \textit{real} or \textit{generated/fake} \cite{qgan_02}.
\end{enumerate}

However, the following assumptions are made in this framework:
\begin{enumerate}
    \item First, it is assumed that both the quantum information processors $G$ and $D$ have enough \textit{capacity} \cite{eqgan} or \textit{expressibility} \cite{expressibility} such that they can approximate any arbitrary function or transformation using their parameters $\overrightarrow{\theta_{g}}$ and $\overrightarrow{\theta_{d}}$ respectively.

    \item Second, it is assumed that an efficient optimization scheme exists that can drive the algorithm to 
    \textit{converge} to the unique Nash equilibrium.

\end{enumerate}
The highlighted assumptions are detailed in Appendix \ref{appendix:qgan_caveats}.

\begin{figure*}
    \centering
    
    \includegraphics[width=\linewidth]{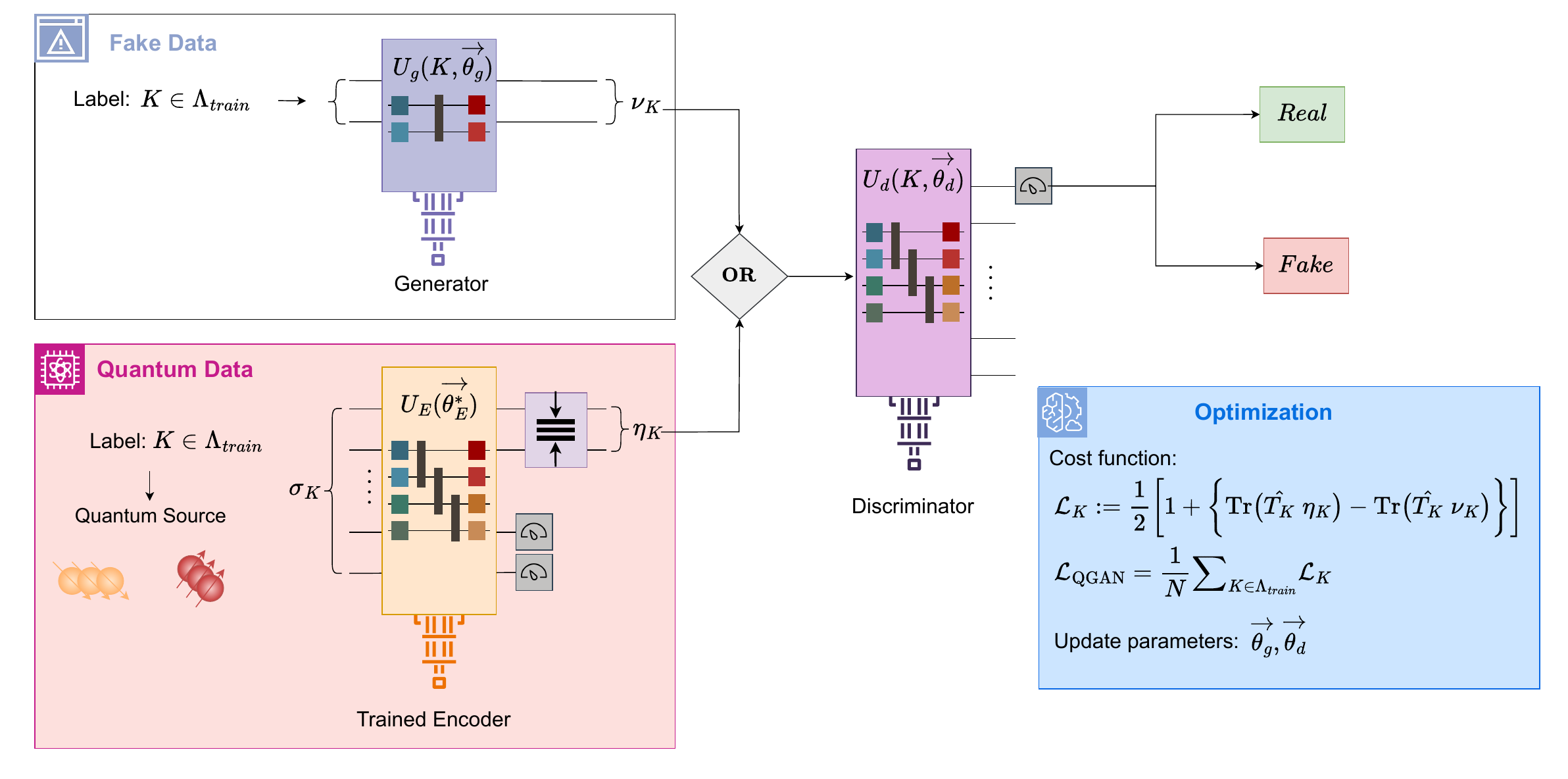}
    \caption{\textbf{QGAA adversarial learning stage} $-$ The states $\{ \eta_{k} \}$ obtained from the \textit{trained encoder} $U_{E} (\overrightarrow{\theta_{E}^{*}})$ of the QAE are the \textit{real data} for training the QGAN.
    The training label set, $\Lambda_{train}$, is the set of all labels whose corresponding $\{ \sigma_{K} \}$ are used to train the QGAA.
    The \textit{fake data} $\{ \nu_{K} \}$, conditioned on $K \in \Lambda_{train}$, is generated by the generator $U_{g} (K, \overrightarrow{\theta_{g}})$. In a training iteration, the discriminator $U_{d} (K, \overrightarrow{\theta_{d}})$ estimates the likelihood of the input state ($\eta_{K}$ or $\nu_{K}$) supplied to it being \textit{real}.  
    The objective of the discriminator is to optimize $\overrightarrow{\theta_{d}}$ to correctly estimate $\eta_{k}$ as \textit{real} and $\nu_{k}$ as \textit{fake} by maximizing the cost function $\mathcal{L}_{\text{QGAN}}$. Whereas, the objective of the generator is to optimize $\overrightarrow{\theta_{g}}$ to generate $\nu_{K}$ whose likelihood of being classified as \textit{fake} is minimized. The adversarial training implements these objectives via the min-max optimization of the cost function $\mathcal{L}_{QGAN}$ by the generator and the discriminator respectively. The Nash equilibrium of the adversarial training is reached when the discriminator can no longer distinguish between $\eta_{k}$ and $\nu_{k}$. At this point, the latent space representation of the \textit{trained encoder} has been learned since $\nu_{K}(\overrightarrow{\theta_{g}^{*}}) = \eta_{K}$
    $\forall \; K \in \Lambda_{train}$ for some optimal parameters $\overrightarrow{\theta_{g}^{*}}$. 
    If $\Lambda_{train}$ is chosen to be a representative sample of the larger set $\Lambda$, then it is expected that $\nu_{K}(\overrightarrow{\theta_{g}^{*}}) = \eta_{K}$
    $\forall \; K \in \Lambda$ as well.}
    \label{fig:qgaa_training_state}
\end{figure*}

\subsubsection{Conditional QGAN}
As an augmentation over QGAN, the quantum source, $R$, 
can have an additional description such that a label, $K \in \Lambda$,
can be supplied to it as an input, conditioning 
$R \equiv R(K)$
with the density matrix representation $\sigma_{K}$. Similar to the notation defined in Section \ref{sec:background_qae}, $\Lambda := \{ K \}$ denotes the set of all valid labels that can be supplied to $R$ to produce the set of quantum states $\Gamma := \{ \sigma_{K} \}$.
The QGAN formalism described so far can also be extended to train the generator, $G$, to generate a \textit{fake state}, $\rho_{K}$, conditioned on $K$, such that, $\rho_{K} = \sigma_{K} \; \forall \; K \in \Lambda$ at the \text{Nash equilibrium} of the quantum adversarial game \cite{qgan_02}. 

Similar to the QGAN training, the quantum adversarial game for the conditional QGAN is as follows:
\begin{enumerate}
    
    \item The parameters $\overrightarrow{\theta_{g}}$ and $\overrightarrow{\theta_{d}}$ of the generator, $G$, and the discriminator, $D$, respectively are initialized.
    Conditioned on the label, $K$, the \text{real source}, $R \equiv R( K)$, produces, $\sigma_{K}$, and the initial strategies of $G$ and $D$ can be described by $\rho_{K}$ and $\hat{T}_{K}$ respectively where,
    \begin{subequations}

        \begin{equation}
        \label{eq:conditional_qgan_gen_state_k}
            \rho_{K} \equiv G (K, \overrightarrow{\theta_{g}})
            \text{ and }
            \hat{T}_{K} \equiv \hat{T}(K, \overrightarrow{\theta_{d}}).
        \end{equation}

Based on Equation \ref{eq:qgan_cost},
the probability of distinguishing between $\sigma_{K}$ and $\rho_{K}$ using the discriminating strategy $\hat{T}_{K}$ is evaluated as,
    
        \begin{equation}
        \label{eq:conditional_qgan_cost_k}
        \mathcal{L}_{K} := \frac{1}{2} \bigg[ 1 + \bigg\{ \text{Tr}  \big( \hat{T_{K}} \; \sigma_{K}  \big) - \text{Tr} \big( \hat{T_{K}} \; \rho_{K} \big)  \bigg\} \bigg].
    \end{equation}

The cost function, $\mathcal{L}_{\text{QGAN}}$, for the adversarial training 
is defined as,
    \begin{equation}
    \label{eq:conditional_qgan_cost_model}
        \mathcal{L}_{\text{QGAN}} = \frac{1}{N} {\sum}_{K \in \Lambda} \mathcal{L}_{K},
    \end{equation}
    the average of the probabilities of discrimination corresponding to different labels, $K \in \Lambda$ \cite{qgan_02}.

    \end{subequations}

    \item The strategy of $D$ is to optimize $\overrightarrow{\theta_{d}}$ to produce $\hat{T}^{\prime}_{K}$ that maximizes $\mathcal{L}_{\text{QGAN}}$ keeping the parameters of $G$ fixed.

    \item The strategy of $G$ is to optimize $\overrightarrow{\theta_{g}}$ to generate $\rho^{\prime}_{K}$ such that Tr$\{ \hat{T}^{\prime}_{K} \rho^{\prime}_{K} \}$ gets closer to Tr$\{ \hat{T}^{\prime}_{K} \sigma_{K} \}$ and minimizes $\mathcal{L}_{\text{QGAN}}$ while keeping the parameters of $D$ fixed.

    \item Subsequent iterations of Step 2 and Step 3 correspond to the max-min optimization of $\mathcal{L}_{\text{QGAN}}$ in Equation \ref{eq:conditional_qgan_cost_model}. 
    A \text{unique Nash equilibrium} exists for some optimal parameters $\overrightarrow{\theta_{g}^{*}}$ such that $\rho_{K}(\overrightarrow{\theta_{g}^{*}}) = \sigma_{K}$ 
    $\forall \; K \in \Lambda$ \cite{qgan_01, qgan_02}.  
    
    \item Upon convergence to the unique fixed-point of the game, $D$ \textit{either} distinguishes between $\sigma_{K}$ and $\rho_{K}(\overrightarrow{\theta_{g}^{*}})$ $\forall \; K \in \Lambda$ as good as a fair-coin toss \cite{qgan_01} \textit{or} identically concludes that the state is \textit{real} no matter whether the input state is \textit{real} or \textit{fake} (i.e. generated) \cite{qgan_02}.

\end{enumerate}

This framework automatically requires $R$ to be controllable with respect to $K$. Such a structure also necessitates that $G$ is equipped with a suitable feature map that can encode any label, $K \in \Lambda$, and that the space of density matrices $\{ \rho_{K}(\overrightarrow{\theta_{g}}) \}$ contains the \textit{unique fixed-point} $\rho_{K} = \sigma_{K} \; \forall \; K \in \Lambda$ for the optimal parameters $\overrightarrow{\theta_{g}^{*}}$. For quantum generative tasks such as those described in Sections \ref{sec:example_1} and \ref{sec:example_2}, the encoded label is a continuous variable. In such cases, a representative sample $\Lambda_{train} \in \Lambda$ can be chosen whose corresponding states $\{ \sigma_{K} \}_{train}$ can be used for the adversarial training.  \\

Section \ref{sec:adversarial_formalism} describes how this formalism of conditional QGAN can be leveraged to confer generative capability to the traditional QAE that is designed to only perform quantum state compression and reconstruction.
\section{Adversarial Formalism for learning Quantum Latent Space}

\label{sec:adversarial_formalism}

The work detailed in this section draws inspiration from the classical Adversarial Autoencoder \cite{adversarial_autoencoder}, where, adversarial training is used to regularize the latent space of a classical Autoencoder. Appendix \ref{appendix:adversarial_autoencoder} describes this classical machine learning architecture. 

In this work, we propose 
using  quantum adversarial learning to model the latent space representation, $\{ \eta_{K} \}$, of a trained QAE.
$\{ \eta_{K} \}$ 
essentially captures the core features of the original quantum data $\Gamma := \{ \sigma_{K} \}$,
and the adversarial learning approach enables direct access to $\{ \eta_{K} \}$.

With the \textit{trained encoder} of a QAE as the real source $R$, we demonstrate that, under the assumptions described in Section \ref{sec:QGAN}, the quantum generative adversarial learning approach can be used to directly generate quantum latent states, $\{ \nu_{K} = \eta_{K} \}$, corresponding to the labels, $K \in \Lambda$. 
Then the generated latent states, $\{ \nu_{K} \}$, can be supplied to the corresponding \textit{trained decoder}
to generate the reconstructed quantum states, $\{ \xi_{K} 
 = \rho_{K} \}$. If the trained QAE performs perfect reconstruction, i.e., $\{ \rho_{K} 
 = \sigma_{K} \}$, then $\{ \xi_{K} 
 = \sigma_{K} \}$.

The following quantum adversarial game to achieve this is detailed in Algorithm \ref{alg:fully_quantum_adversarial_autoencoder} and depicted in Figure
\ref{fig:qgaa_training_state}: 
\begin{enumerate}
    
    \item The parameters $\overrightarrow{\theta_{g}}$ and $\overrightarrow{\theta_{d}}$ of the generator, $G$, and the discriminator, $D$, respectively are initialized, and a 
    representative sample
    of training labels, $\Lambda_{train} := \{ K \}_{train} \in \Lambda$ is defined, where, $N = |\Lambda_{train}|$, is the size of the training label set.
    The \text{real source}, $R \equiv R( K)$, comprises the \textit{trained encoder}, $U_{E}(\overrightarrow{\theta^{*}_{E}})$, which can produce the compressed representation $\eta_{K}$, corresponding to the \textit{real state}, $\sigma_{K}$,  
    conditioned on the label $K$. 
    Similarly, $G$ can generate a \textit{fake state}, $\nu_{K}$, conditioned on the label, $K$, and the initial strategy of $D$ can be described by the corresponding positive semi-definite operator $\hat{T}_{K}$,
    \begin{subequations}

        \begin{equation}
        \label{eq:qgan_gen_state_k}
            \nu_{K} \equiv G (K, \overrightarrow{\theta_{g}})
            \text{ and }
            \hat{T}_{K} \equiv \hat{T}(K, \overrightarrow{\theta_{d}}).
        \end{equation}

Similar to Equation \ref{eq:conditional_qgan_cost_k},
the probability of distinguishing between $\eta_{K}$ and $\nu_{K}$ using the discriminating strategy $\hat{T}_{K}$ is evaluated as,

        \begin{equation}
        \label{eq:qgan_cost_k}
        \mathcal{L}_{K} := \frac{1}{2} \bigg[ 1 + \bigg\{ \text{Tr}  \big( \hat{T_{K}} \; \eta_{K}  \big) - \text{Tr} \big( \hat{T_{K}} \; \nu_{K} \big)  \bigg\} \bigg].
    \end{equation}

Similar to Equation \ref{eq:conditional_qgan_cost_model}, the cost function, $\mathcal{L}_{\text{QGAN}}$, for the adversarial training 
is defined as,
    \begin{equation}
    \label{eq:qgan_cost_model}
        \mathcal{L}_{\text{QGAN}} = \frac{1}{N} {\sum}_{K \in \Lambda_{train}} \mathcal{L}_{K},
    \end{equation}
    the average of the probabilities of discrimination corresponding to different labels, $K \in \Lambda_{train}$ \cite{qgan_02}.

    \end{subequations}

    \item The strategy of $D$ is to optimize $\overrightarrow{\theta_{d}}$ to produce $\hat{T}^{\prime}_{K}$ that maximizes $\mathcal{L}_{\text{QGAN}}$ keeping the parameters of $G$ fixed.

    \item The strategy of $G$ is to optimize $\overrightarrow{\theta_{g}}$ to generate $\nu^{\prime}_{K}$ such that Tr$\{ \hat{T}^{\prime}_{K} \nu^{\prime}_{K} \}$ gets closer to Tr$\{ \hat{T}^{\prime}_{K} \eta_{K} \}$ and minimizes $\mathcal{L}_{\text{QGAN}}$ while keeping the parameters of $D$ fixed.

    \item Subsequent iterations of Step 2 and Step 3 correspond to the max-min optimization of $\mathcal{L}_{\text{QGAN}}$ in Equation \ref{eq:qgan_cost_model}. 
    A \text{unique Nash equilibrium} exists for some optimal parameters $\overrightarrow{\theta_{g}^{*}}$ such that $\nu_{K}(\overrightarrow{\theta_{g}^{*}}) = \eta_{K}$
    $\forall \; K \in \Lambda_{train}$ \cite{qgan_01, qgan_02}.

    \item Upon convergence to the unique fixed-point of the game, $D$ \textit{either} distinguishes between $\eta_{K}$ and $\nu_{K}(\overrightarrow{\theta_{g}^{*}})$ $\forall \; K \in \Lambda_{train}$ as good as a fair-coin toss \cite{qgan_01} \textit{or} identically concludes that the state is \textit{real} no matter whether the input state is \textit{real} or \textit{fake} (i.e. generated) \cite{qgan_02}.
\end{enumerate}

\SetKwComment{Comment}{/* }{ */}
\begin{algorithm}[!t]
    \caption{Adversarial training for learning the latent space of a Quantum Autoencoder.} 
    
    \label{alg:fully_quantum_adversarial_autoencoder}
    
    \textbf{Input Quantum} \KwData{$\{ \eta_{K} \}$}
    
    \KwResult{$\nu_{K} = \eta_{K} \; \forall \; K \in \Lambda$}

    \vspace{0.5em}

    $U_{E}^{*} \equiv U_{E}(\overrightarrow{\theta^{*}_{E}})$ \;

    \vspace{0.5em}

    The Generator 
    $U_{g}(K, \overrightarrow{\theta_{g}})$ generates $\nu_{K}(\overrightarrow{\theta_{g}})$\;

    \vspace{0.5em}
    
    {The Discriminator $U_{d} \equiv U_{d}(K, \overrightarrow{\theta_{d}})$ evaluates the probabilities $\text{Prob} \big(+1|\eta_{K})$ and $\text{Prob} \big(+1|\nu_{K})$ corresponding to the probe qubit determining the input state ($\eta_{K}$ or $\nu_{k}$) to be \textit{real} (+1)} \;

    \vspace{0.5em}
    
    Initialize the parameters $\overrightarrow{\theta_{g}}$ and $\overrightarrow{\theta_{d}}$,
    \textit{iter} $= 1$\;

    Define maximum number of iterations: \textit{max}{\textunderscore}\textit{iter} \;

    \vspace{0.5em}
     
    \While{\textit{iter $\leq$ max{\textunderscore}iter}}
    {               

        



    \While{$K \in \Lambda_{train}$}
    {
        \vspace{0.5em}

        $\eta_{K} := \text{Tr}_{trash} \bigg\{ U_{E}^{*} \; \sigma_{K} \;  {U_{E}^{*}}^{\dagger} \bigg\}$

        \vspace{0.5em}

        $\braket{Z_{real}} :=
        \text{Tr} \bigg[ \bigg\{ U_{d} \big( \eta_{K} \otimes \ket{0}\bra{0}_{probe} \big) U_{d}^{\dagger} \bigg\} Z_{probe}  \bigg]$

        \vspace{0.5em}

        $\therefore \text{Prob} \;\big( \text{probe}=+1 \; \big| \; \eta_{K} \big) = \frac{1 + \braket{Z_{real}}}{2}$

        \vspace{0.5em}

        $\braket{Z_{fake}} :=
        \text{Tr} \bigg[ \bigg\{ U_{d} \big( \nu_{K} \otimes \ket{0}\bra{0}_{probe} \big) U_{d}^{\dagger} \bigg\} Z_{probe}  \bigg]$

        \vspace{0.5em}

        $\therefore \text{Prob} \; \big( \text{probe}=+1 \; \big| \; \nu_{K} \big) = \frac{1 + \braket{Z_{fake}}}{2}$
        
        \vspace{0.5em}
        
        \textbf{Success  Probability as cost function}:
        \hfill \break
        $\mathcal{L}_{K} := 
        \frac{1}{2} \bigg[ 1 + \big\{ \text{P} ( +1 | \eta_{K} ) -
        \text{P} ( +1 | \nu_{K} ) \big\} \bigg]$
    }

    \vspace{0.5em}
    $\mathcal{L}_{\text{QGAN}} = \frac{1}{N} {\sum}_{K \in \Lambda_{train}} \mathcal{L}_{K}$
    
    \vspace{0.5em}
    $\longrightarrow$ \textbf{Discriminator's Strategy:}\hspace{5em}
    Optimize and Update $\overrightarrow{\theta_{d}}$ to maximize $\mathcal{L}_{\text{QGAN}}$ \;
    
    \vspace{0.5em}
    \textbf{Compute} $\mathcal{L}_{\text{QGAN}}$ \textbf{using updated} $\overrightarrow{\theta_{d}}$ \;

    \vspace{0.5em}
    $\longrightarrow$ \textbf{Generator's Strategy:} \hspace{5em}
    Optimize and Update $\overrightarrow{\theta_{g}}$ to minimize $\mathcal{L}_{\text{QGAN}}$ \;

    \vspace{0.5em}
    
    $iter \; += \; 1$

    }
\end{algorithm}

\begin{figure*}[!tbh]
    \centering
  
    \includegraphics[width =0.82\linewidth]{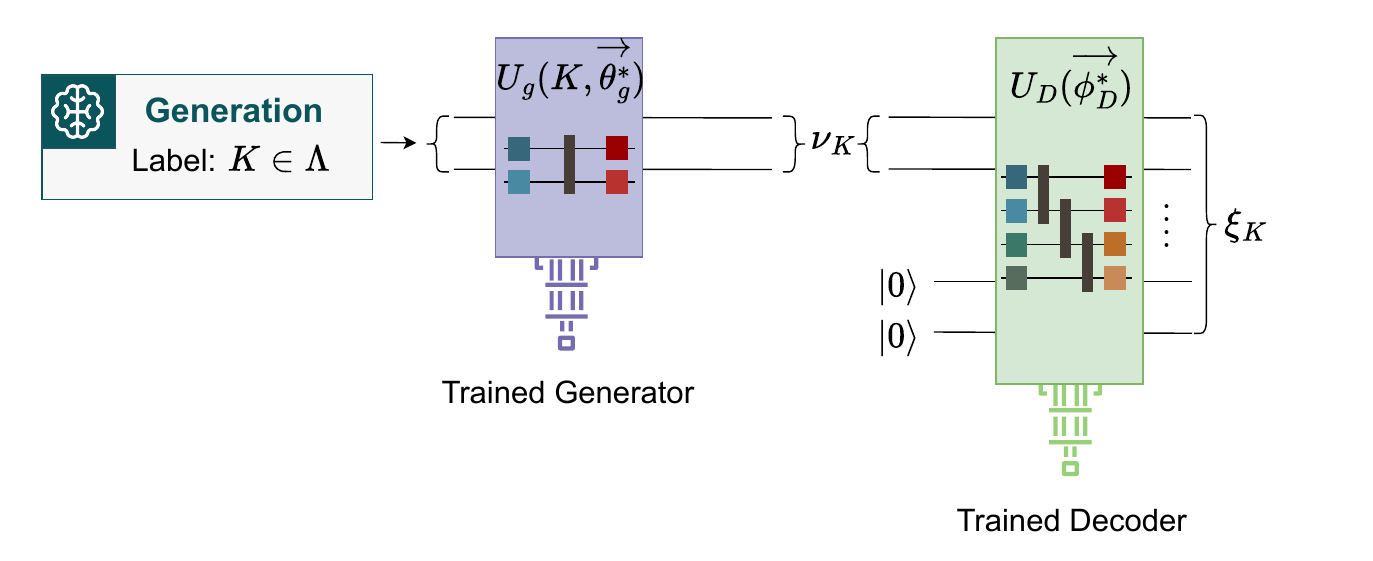}
    \caption{\textbf{QGAA generation stage}$ - $ Upon learning the latent space representation, the \textit{trained generator} $U_{g} (K, \overrightarrow{\theta_{g}^{*}})$ is capable of directly generating the latent state $\nu_{K} = \eta_{K}$ corresponding to any label $K \in \Lambda$. \hfill \newline
    The generated latent state $\nu_{K}$ is fed into the \textit{trained decoder} $U_{D}(\overrightarrow{\theta_{D}^{*}})$ of the QAE to reconstruct $\xi_{K} = \rho_{K}$. If the trained QAE performs perfect reconstruction, i.e., $\{ \rho_{K} 
    = \sigma_{K} \}$, then $\{ \xi_{K} 
    = \sigma_{K} \}$. Thus, the quantum adversarial protocol is able to give generative capabilities to a QAE by learning the representation of its latent space.
    }
    \label{fig:model_generation}
\end{figure*}

After the model has been trained to converge to its Nash equilibrium, $G$ is capable of directly generating latent states corresponding to any label, $K \in \Lambda_{train}$. Since $\Lambda_{train}$ is chosen to be a representative sample of the larger set $\Lambda$, it is expected that $\nu_{K}(\overrightarrow{\theta_{g}^{*}}) = \eta_{K}$ $\forall \; K \in \Lambda$ as well.
The generated latent state, $\nu_{K}$, can be fed into the \textit{trained decoder}, $U_{D}(\overrightarrow{\theta^{*}_{D}})$, of the QAE to reconstruct $\xi_{K} = \rho_{K}$ as depicted in Figure \ref{fig:model_generation}. Further, in case of perfect reconstruction capabilities of the trained QAE, $\xi_{K} = \sigma_{K}$. Thus, the quantum adversarial protocol is able to give generative capabilities to a QAE by learning the representation of its latent space. 

Through informative applications in the following Sections \ref{sec:example_1} and \ref{sec:example_2}, we investigate this formalism to learn the latent space representations of trained QAEs and generate quantum data directly from the latent spaces.
We show that the proposed adversarial learning scheme
is capable of generating new quantum states (corresponding to labels outside the training dataset) exhibiting a particular property, such as entanglement and ground states of a parameterized molecular Hamiltonian.
\section{Application Example $\mathbf{1}$: 
\newline
Learning latent representation of Entangled States}

\label{sec:example_1}

This section demonstrates the application of the adversarial formalism to learn the latent space representation of entangled states. The following 2-qubit entangled state is prepared conditioned on the label information $K = ( k_{0}, k_{1} )$:
\begin{subequations}
\begin{equation}
    \label{eq:entangled_input_state}
        \ket{\psi_{K}} = cos \bigg( \frac{k_{0}}{2} \bigg) \ket{00} + e^{ik_{1}}sin \bigg( \frac{k_{0}}{2} \bigg) \ket{11},
\end{equation}

\begin{equation}
    \label{eq:set_entangled_input_state}
        \sigma_{K} = \ket{\psi_{K}}\bra{\psi_{K}}.
\end{equation}
\end{subequations}
Here, the label, $K$, is further comprised of sub-labels, $k_{0}$ and $k_{1}$, that characterize $\sigma_{K}$. \\
 
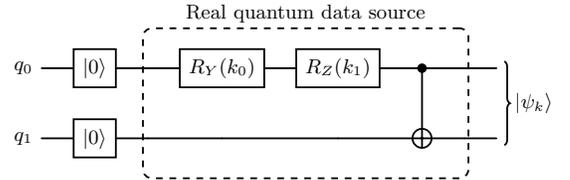
\begin{figure}[h]
    \centering 
    \scalebox{0.85}{
    \begin{quantikz}
        \lstick{$q_{0}$} &\gate{\ket{0}}
        &
        \gategroup[2, steps=4,style={dashed, inner sep=6pt, rounded corners, xshift=+0.25cm}]{Real quantum data source}
        &   \gate{R_Y(k_{0})} & \gate{R_Z(k_{1})} &\ctrl{1}&&\rstick[2]{$\ket{\psi_{k}}$}\\
        \lstick{$q_{1}$}&\gate{\ket{0}}&&&&\targ{}&&
    \end{quantikz}

    }
    \caption{Quantum circuit for generating 2-qubit parameterized entangled states $\ket{\psi_{k}}$ conditioned on the label information $K = (k_{0}, k_{1}) \in \Lambda$ supplied as input.}
    \label{fig:entangled_states_circuit}
\end{figure}

$\ket{\psi_{K}}$ is created by first preparing the single qubit state
$\ket{\chi_{K}}$ 
conditioned on the sub-labels $k_{0}$ and $k_{1}$: 
\begin{subequations}
    \begin{multline}
    \label{eq:entangled_input_space}
        \ket{\chi_{K}} = RZ \big( k_{1} \big) \; RY \big( k_{0} \big) \; \ket{0} \\
            = cos \bigg( \frac{k_{0}}{2} \bigg) \ket{0} + e^{ik_{1}}sin \bigg( \frac{k_{0}}{2} \bigg) \ket{1}.
    \end{multline}

Applying the CX gate on another qubit $q_{1}$ (in the state $\ket{0}$) with qubit $q_{0}$ (in the state $\ket{\chi_{K}}$) as the control, generates the 2-qubit entangled state $\ket{\psi_{K}}$ as depicted in Figure \ref{fig:entangled_states_circuit}: 
\begin{equation}
    \label{eq:psi_cx_chi}
       \ket{\psi_{K}} = \text{CX} \ket{\chi_{K}} \ket{0}.
\end{equation}
\end{subequations}

Therefore, the set $\big\{ \ket{\chi_{K}} \big\}$ is a 1-qubit latent space representation of $\big\{ \sigma_{K} \big\}$
since the reversible operation of a CX gate on $\ket{{\psi_{K}}}$ can ``compress'' it to a corresponding 1-qubit representation, $\ket{\chi_{K}}$.
It is worthwhile to note that this 1-qubit latent space representation of $\big\{ \sigma_{K} \big\}$ is not unique. 
A QAE upon iterative training can converge to a 1-qubit latent representation different than the one in Equation \ref{eq:entangled_input_space}.

\begin{figure*}[!thb]

\centering

\begin{subfigure}[h]{0.45\textwidth}
    \centering
    \scalebox{0.85}{
    \begin{quantikz}
        \lstick[2]{$\sigma_{k} = \ket{\psi_{k}}\bra{\psi_{k}}$} &
        \gategroup[2, steps=4,style={dashed, inner sep=6pt, rounded corners, xshift=+0.2cm}]{Encoder: $U_{E} (\overrightarrow{\theta_{E}})$}
        & \gate{R_Y(\theta_0)} &
        \ctrl{1} & \gate{R_Y(\theta_2)} & 
        & \rstick{$\eta_{k}$} \\
        &&\gate{R_Y(\theta_1)}& \control{} & \gate{R_Y(\theta_3)} &&  \rstick{\textit{trash}}
    \end{quantikz}
   
    }
   \caption{Compression of $\sigma_{k}$ to $\eta_{k}$.}

    \label{fig:qae_encoder_ansatz}
\end{subfigure}
\hfill
\begin{subfigure}[h]{0.45\textwidth}
    \centering
    \scalebox{0.85}{
    \begin{quantikz}
        \lstick[1]{$\eta_{k}$}&&
        \gategroup[2, steps=4,style={dashed, inner sep=6pt, rounded corners, xshift=+0.3cm}]{Decoder: $U_{D} (\overrightarrow{\phi_{D}})$}
        &  \gate{R_Y(\phi_0)} & \ctrl{1} & \gate{R_Y(\phi_2)} && \rstick[2]{$\rho_{k}$} \\
        \lstick[1]{\textit{auxiliary }}&\gate{\ket{0}}
        &&\gate{R_Y(\phi_1)}& \control{} & \gate{R_Y(\phi_3)} && 
    \end{quantikz}
    
    }
   \caption{Reconstruction of $\rho_{k}$ from $\eta_{k}$.}

    \label{fig:qae_decoder_ansatz}
\end{subfigure}

    \caption{Parameterized quantum circuits of (a) the encoder to compress $2$-qubit entangled states $\{ \sigma_{k} \}$ to a $1$-qubit representation $\{ \eta_{k} \}$ by discarding the processed \textit{trash qubit}, and (b) the decoder to reconstruct of 2-qubit entangled states $\{ \rho_{k} \}$ from $\{ \eta_{k} \}$. The parameters $\overrightarrow{\theta_{E}}$ and $\overrightarrow{\phi_{D}}$ of the encoder and the decoder respectively are optimized during the QAE training to achieve $\mathcal{F}(\sigma_{k}, \rho_{k}) = 1$.}

    \label{fig:qae_ansatz}
\end{figure*}
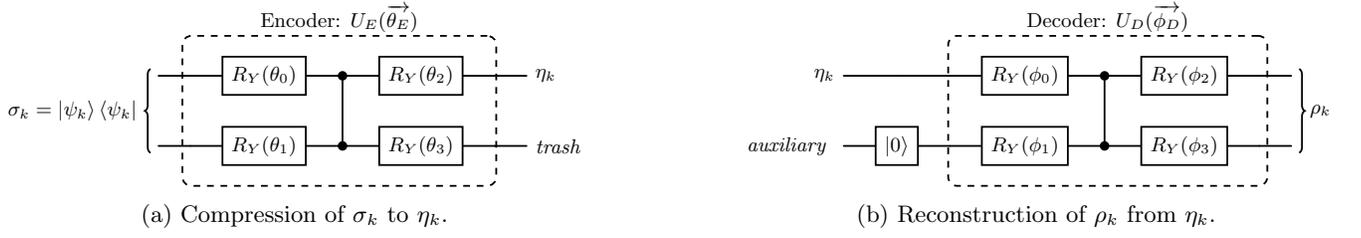

\begin{figure}[!h]
    \centering
   \includegraphics[scale=0.35]{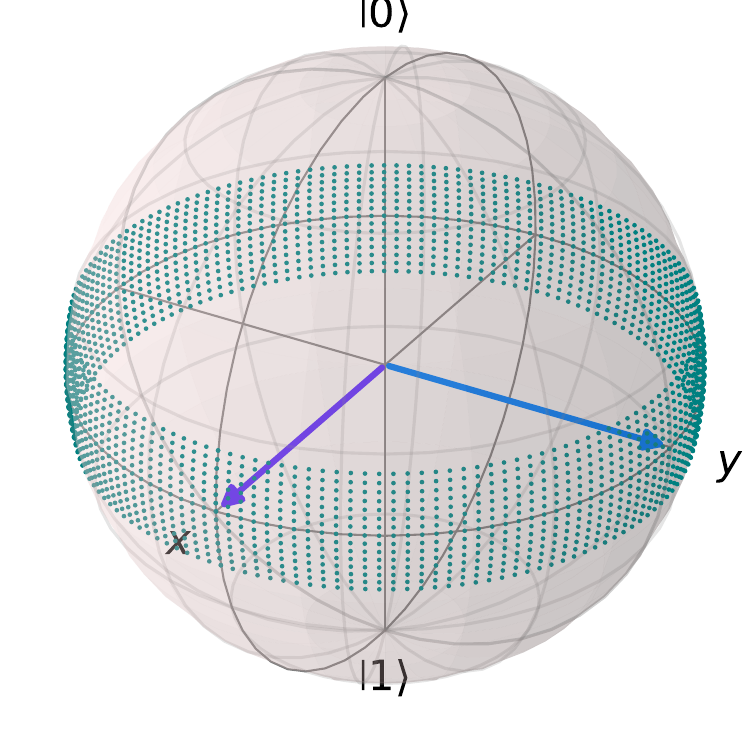}
   \caption{1-qubit pure states  $\big\{ \ket{\chi_{K}} \big\}$ on the Bloch sphere corresponding to $k_{0} \in  \bigg[ \frac{\pi}{2} - 0.06\pi, \frac{\pi}{2} + 0.06\pi \bigg]$ and $k_{1} \in [0, 2\pi)$. The two states marked by Bloch vectors correspond to $K = ( 0.5\pi, 0 )$ and $K = ( 0.5\pi, 0.5\pi )$.}    \label{fig:entangled_states_input_space}
\end{figure}

For ease of visualization in the following discussion, we consider the set of training labels, $\Lambda_{train}$, to be the set of labels $\{ K \}_{train}$, whose corresponding $\{ \sigma_{K} = \ket{\psi_{K}}\bra{\psi_{K}}\}$ comprises all the states  with entanglement entropy greater than $97\%$. The corresponding $\ket{\chi_{K}}$ are depicted on the Bloch sphere in Figure \ref{fig:entangled_states_input_space}. The entanglement entropy of $\{ \ket{\psi_{K}} \}$ corresponding to $k_{0} = 0.5\pi \pm 0.06\pi$ and $k_{0} = 0.5\pi$ is 0.97 and 1.00 respectively.

\subsection{Training the QAE for state compression}

\begin{figure}[!t]
    \centering
    \includegraphics[width=0.35\textwidth]{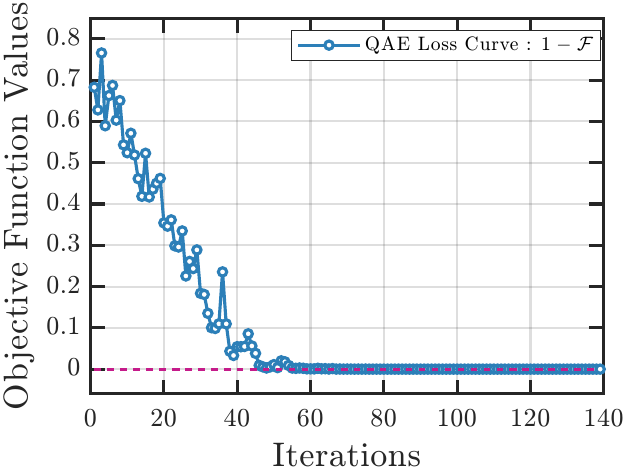}
                   
    \caption{Plot depicting the minimization of the QAE cost function (Equation \ref{eq:qae_cost_function}) for optimal compression (from 2 qubits to 1 qubit) and reconstruction of entangled states (Equation \ref{eq:entangled_input_state}).} 
    \label{fig:Entanglement_QAE_convergence}
\end{figure}

The QAE is optimized according to Section \ref{sec:background_qae} to compress the 2-qubit entangled states $\sigma_{K} = \ket{\psi_{K}}\bra{\psi_{K}}$ to a 1-qubit latent representation $\eta_{K}$. Figure \ref{fig:qae_ansatz} depicts the ansatzes used for the encoder and the decoder of the QAE. Figure \ref{fig:Entanglement_QAE_convergence} depicts the decreasing loss function of the QAE upon training using the COBYLA optimizer.

\begin{figure}[!h]
\centering
\includegraphics[scale=0.35]{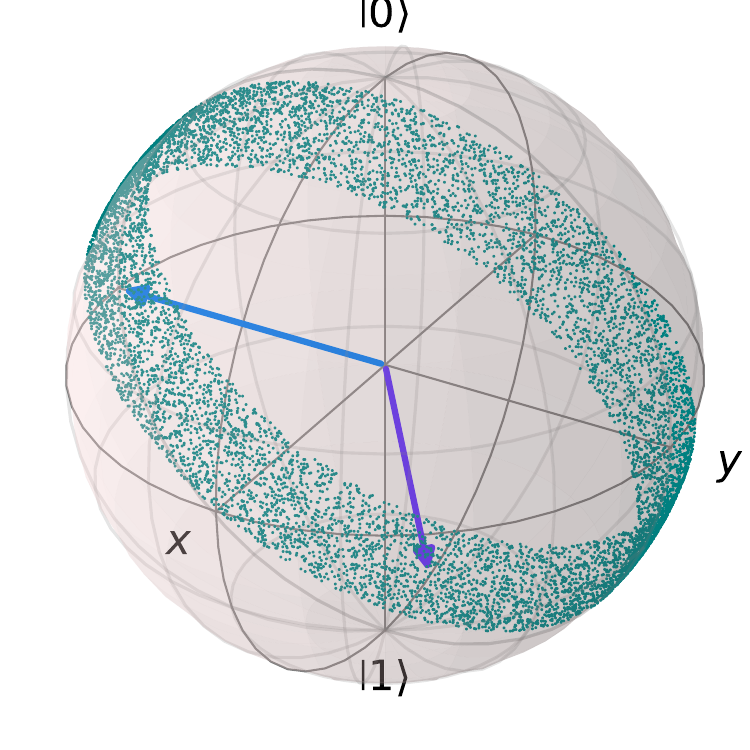}
        
\caption{1-qubit latent space $\big\{ \ket{\gamma_{K}} \big\}$ generated by the \textit{trained encoder} upon compression of 2-qubit entangled states $\{ \sigma_{K} \}$. The two states marked by Bloch vectors correspond to $K = \{ 0.5\pi, 0 \}$ and $K = \{ 0.5\pi, 0.5\pi \}$.}

\label{fig:qae_latent_space}

\end{figure}

The QAE achieves perfect reconstruction of the input states $\{ \sigma_{K} \}$ with fidelity of state reconstruction converging to $\mathcal{F}(\sigma_{K}, \rho_{K}) = 1$. The compressed states $\{ \eta_{K} \}$ have a purity equal to 1. Therefore, the states $\{ \eta_{K} \}$ have a pure state representation $\{ \eta_{K} = \ket{\gamma_{K}}\bra{\gamma_{K}} \}$ and lie on the surface of the Bloch sphere as depicted in Figure \ref{fig:qae_latent_space}. 
Following are the optimal parameters of the \textit{trained encoder} and the \textit{trained decoder} that achieve these results:

\begin{subequations}
    \begin{equation}
    \label{eq:qae_trained_encoder}
        \overrightarrow{\theta^{*}_{E}} = \big( 0.5\pi, \pi, 1.18\pi, 0.6\pi \big),
    \end{equation}

    \begin{equation}
    \label{eq:qae_trained_decoder}        \overrightarrow{\phi^{*}_{D}} = \big( 0.4\pi, 0.5\pi, 0, 1.5\pi \big).
\end{equation}

\end{subequations}

\begin{figure*}[!tbh]

    \begin{subfigure}{\textwidth}
        \centering
       \includegraphics[scale=0.7]{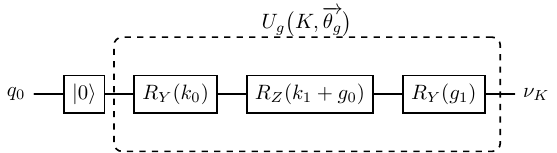}
       \caption{Generator}
    \label{fig:qgan_gen_ansatz}
    \end{subfigure}

    \begin{subfigure}{\textwidth}
        \centering
       
        \includegraphics[scale=0.7]{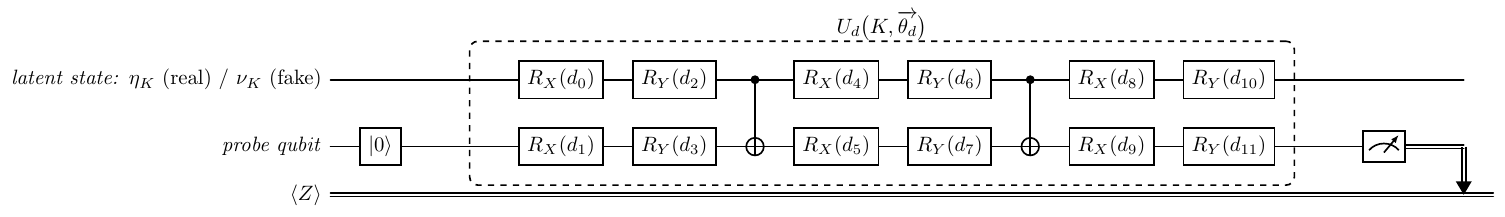}
       \caption{Discriminator}
        \label{fig:qgan_disc_ansatz}
    \end{subfigure}

    \caption{Parameterized quantum circuits of (a) the generator and (b) the discriminator for the adversarial learning of 1-qubit latent representations of entangled states obtained using a trained QAE. The generator generates a 1-qubit state $\nu_{K}$ conditioned on the label information $K = (k_{0}, k_{1})$, and (b) the discriminator evaluates the probability of the input state (either $\eta_{K}$ or $\nu_{K}$) being \textit{real}. The parameters $\overrightarrow{\theta_{g}}$ and $\overrightarrow{\theta_{d}}$ of the generator and the discriminator respectively undergo min-max optimization during the adversarial training.}
    \label{fig:qgan_ansatzes}
    
\end{figure*}

Upon closer numerical inspection of the latent states in Figure \ref{fig:qae_latent_space}, it can be shown that the \textit{trained encoder} $U_E(\overrightarrow{\theta^{*}_{E}})$ effectively implements the following transformation on the state $\ket{\chi_{K}}$ in Equation \ref{eq:entangled_input_space}:
\begin{equation}
    \ket{\gamma_{K}} = RY \big( -(\pi - 0.3) \big) \; RZ \big( \pi \big) \; \ket{\chi_{K}}.
    \label{eq:QAE_state}
\end{equation}

Equation \ref{eq:QAE_state} demonstrates the non-uniqueness of the latent representation since $\eta_{K} = \ket{\gamma_{K}}\bra{\gamma_{K}}$ obtained upon iteratively training the QAE is different than the 1-qubit representation $\ket{\chi_{K}}$ described in Equation \ref{eq:entangled_input_space}.

The two states $\ket{\gamma_{K}}$ corresponding to $k_{0} = \pi/2$ and $k_{1} = \pm \pi/2$ are eigenstates of the effective transformation in Equation \ref{eq:QAE_state}.
For the purpose of visualizing the adversarial learning protocol in the following discussion, it is helpful to keep track of the evolution of one of these eigenstates.

\subsection{Adversarial Learning of QAE Latent Space}

We now implement a QGAN as explained in Section \ref{sec:adversarial_formalism} to learn the representation of the latent space  $\{ \eta_{K} \}$ depicted in Figure \ref{fig:qae_latent_space}. 

Taking insights from the effective transformation of the \textit{trained encoder} (Equation \ref{eq:QAE_state}), we design a suitable single-qubit Euler-decomposition ansatz,
\begin{equation}
\label{eq:qgan_gen_anatz}
    U_{g} \big( K, \overrightarrow{\theta_{g}} \big) = 
    RY \big( g_{1} \big) \;
    RZ \big( g_{0} + k_{1} \big) \;
    RY \big( k_{0} \big),
\end{equation}
as the generator with parameters $\overrightarrow{\theta_{g}} = \big( g_{0}, \; g_{1} \big)$.

Figure \ref{fig:qgan_gen_ansatz} depicts the circuit for the generator that encodes the label $K$ and generates a\textit{ fake state} $\nu_{K} = \ket{\nu_{K}}\bra{\nu_{K}}$.
Figure \ref{fig:qgan_disc_ansatz} depicts the structure of the discriminator ansatz $U_{g} \big( K, \overrightarrow{\theta_{d}} \big)$ with parameters $\overrightarrow{\theta_{d}} =  \big( d_0, ... \; , d_{11} \big)$.

\begin{figure}[!htb]

\centering

\begin{subfigure}[h]{0.2\textwidth}
    \centering
   \includegraphics[scale=0.22]{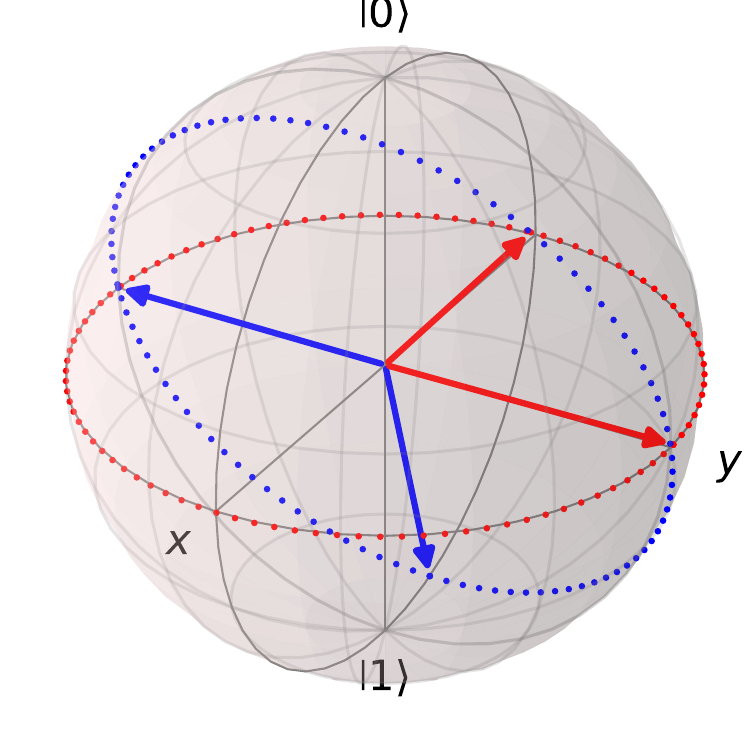}
   \caption{Initial State}    
   \label{fig:entangled_states_qgan_train_0}
\end{subfigure}
\hfill
\begin{subfigure}[h]{0.2\textwidth}
    \centering
   \includegraphics[scale=0.22]{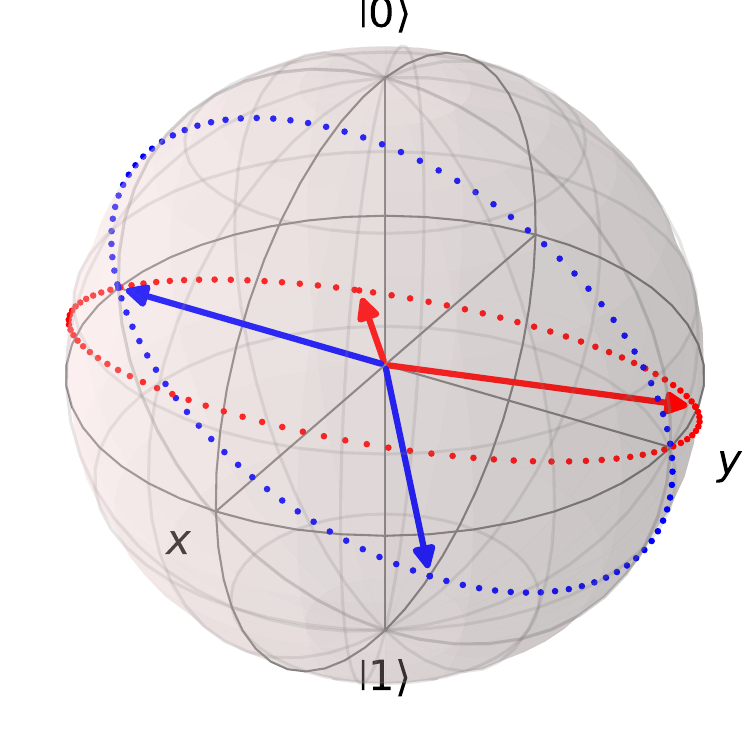}
   \caption{Iteration 50}    
   \label{fig:entangled_states_qgan_train_50}
\end{subfigure}
\hfill
\begin{subfigure}[h]{0.2\textwidth}
    \centering
   \includegraphics[scale=0.22]{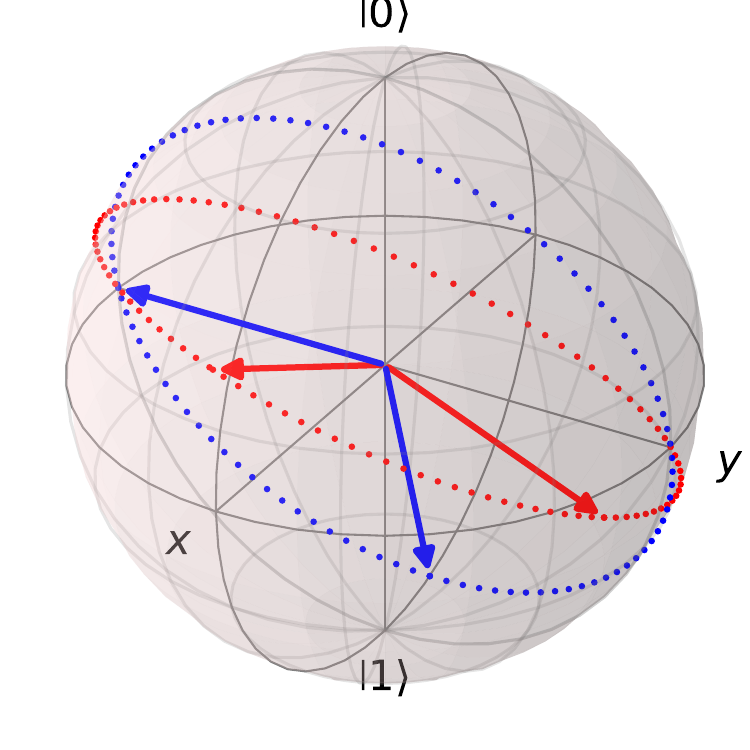}
   \caption{Iteration 75}    
   \label{fig:entangled_states_qgan_train_75}
\end{subfigure}
\hfill
\hfill
\begin{subfigure}[h]{0.2\textwidth}
    \centering
   \includegraphics[scale=0.22]{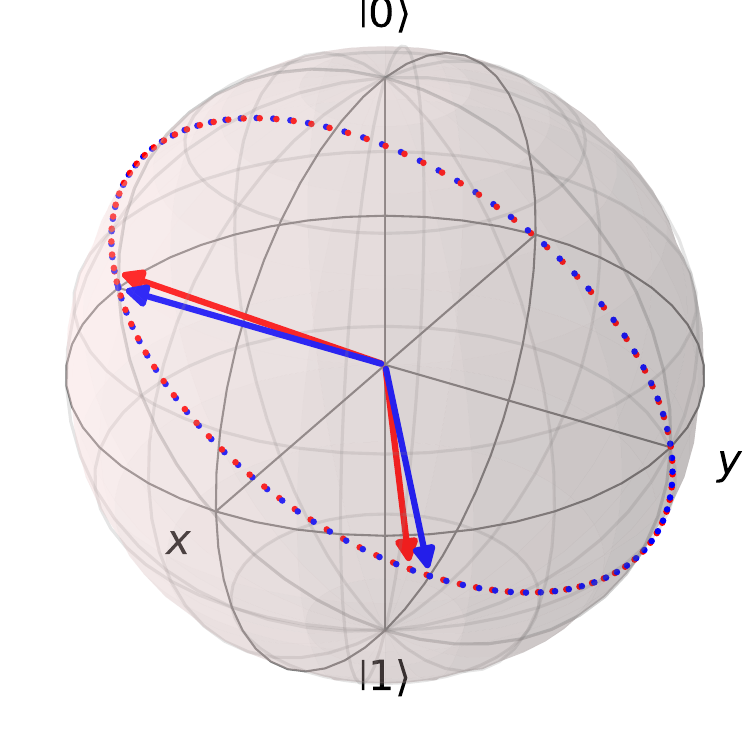}
   \caption{Iteration 110}    
   \label{fig:entangled_states_qgan_train_110}
\end{subfigure}

\caption{Evolution of the learned latent space for $k[0] = \pi/2$ (red points) towards their target latent space (blue points) during adversarial training. Red points show states produced  at a particular iteration of training $\vec{\theta}_g$, while blue points indicate the target states using optimal parameters $\vec{\theta}_g^*$. States corresponding to $k[1] = 0$ and $k[1] = \pi/2$ marked by Bloch vectors for ease of visualization.}
\label{fig:qgan_training_bloch}
\end{figure}

\begin{figure*}[!ptbh]

    \begin{subfigure}{\textwidth}
        \centering
         \includegraphics[width=0.85\linewidth]{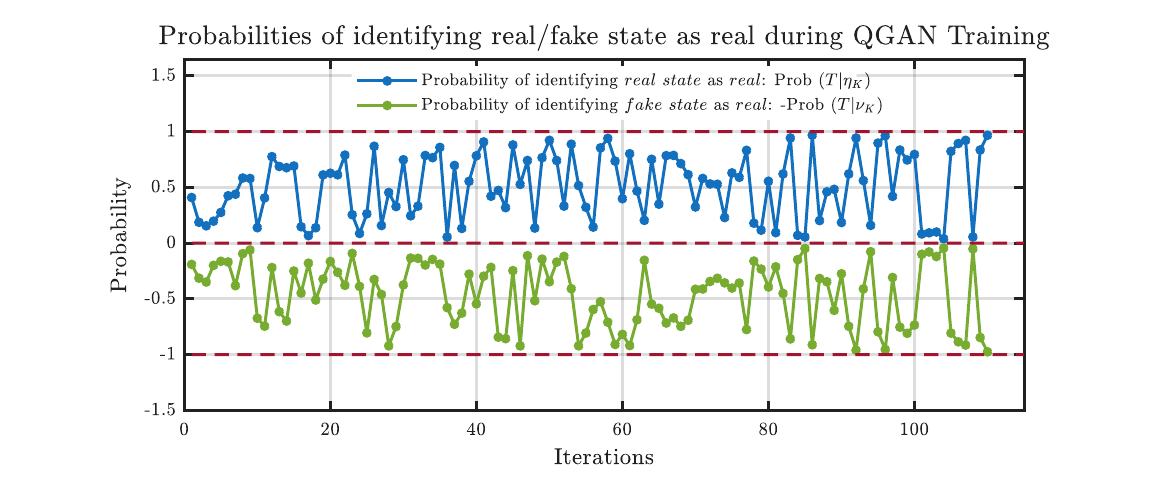}
        \caption{}
        \label{fig:qgan_prob_curves}
    \end{subfigure}
    
    \begin{subfigure}{\textwidth}
        \centering
        \includegraphics[width=0.85\linewidth]{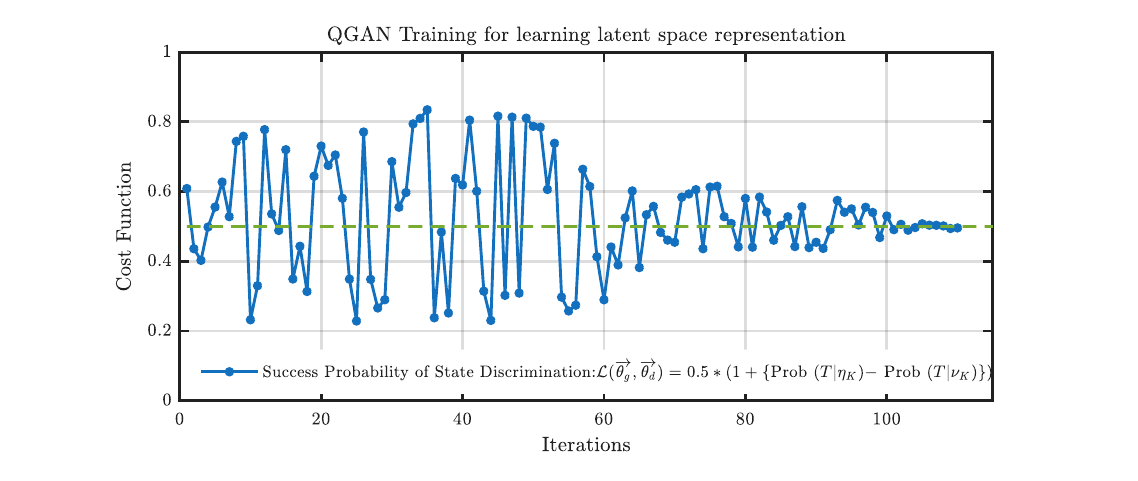}

        \caption{}
        \label{fig:qgan_loss_curves}
    \end{subfigure}

    \begin{subfigure}{\textwidth}
        \centering
        \includegraphics[width=0.85\linewidth]{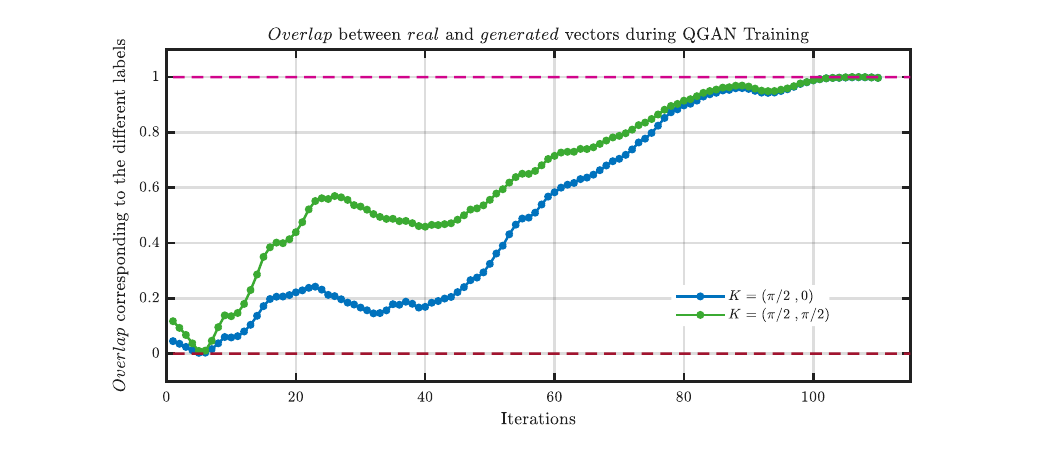}

        \caption{} 
        \label{fig:qgan_vec_overlap}
    \end{subfigure}

    \caption{(a) Probabilities of identifying the real/fake state as \textit{real} during QGAN training. The curves clearly demonstrate adversarial behaviour. At the final 110th iteration of the training, the discriminator identifies both the real and the fake states as \textit{real} with probability very close to $1$. Therefore, at this stage, the generator is able to generate fake states which the discriminator cannot distinguish from real states; (b) This results in the cost function (Equation \ref{eq:qgan_cost_k}) to converge close to $0.5$, implying that the probability of the discriminator to distinguish between real and fake states is as good as a fair-coin toss;
    (c) Further, the $overlap$ between the Bloch vectors of the real $(\eta_k)$ and the generated $(\nu_k)$ state, corresponding to any label $K$, converges to $1$ at the end of the QGAN training. This implies that the generator has indeed learnt the latent representation and can generate latent states $\nu_{K} = \eta_{K}$.}
    \label{fig:qgan_curves}
    
\end{figure*}

\textbf{Note:} Ideally, the discriminator should be equipped with additional qubit(s) that encode the label information $K$ conditioned on which the input state has been created \cite{qgan_02}. This is to ensure that the generator does not \textit{cheat} in the adversarial game by generating an input state of an incorrect label to confuse the discriminator. 
For our examples, we ensure that such a scenario of cheating in the adversarial game does not take place and that the information of the \textit{actual label} is inherently present in the input state and hence implicitly supplied to the discriminator. This assumption reduces the qubit count of the discriminator and makes our instructive examples convenient for demonstrating adversarial learning. 

The \textit{trained encoder} (Figure \ref{fig:qae_encoder_ansatz}) with optimal parameters $\overrightarrow{\theta^{*}_{E}}$ (Equation \ref{eq:qae_trained_encoder}) serves as the source of \textit{real data} $\{ \eta_{K} \}$ for the discriminator, whereas the generator (Figure \ref{fig:qgan_gen_ansatz}) serves as the source of \textit{fake data} $\{ \nu_{K} \}$. A well-trained adversarial strategy should guide the parameters $\overrightarrow{\theta_{g}}$ of the generator   toward their optimal value $\overrightarrow{\theta^{*}_{g}}$ where $\{ \nu_{K} = \eta_{K} \}$. This would mean that the model has been able to learn and directly generate the latent space representation $\{ \eta_{K} \}$ of the \textit{trained} QAE.

Training QGANs is a challenging task and requires careful considerations. 
The initial choice of parameters also greatly determines the trajectory of QGAN training. Good starting points help the optimizer reach the required minimum much easily. After observing the QGAN training behaviour across multiple initializations, we choose the starting parameters as 
$\overrightarrow{\theta_{g}} = (\pi/2, 0)$ and 
$\overrightarrow{\theta_{d}} = \overrightarrow{0}$
with additional noise sampled from a standard normal distribution added to both. This injected noise ensures variability in the initialization, promoting more robust training. We use two Adam optimizers - one each for the generator and the discriminator respectively. The learning rate for the generator is chosen as $10^{-1}$ and that of the discriminator as $10^{-2}$. The evolution of the learning of the latent space can be observed in Figure \ref{fig:qgan_training_bloch}.

Figure \ref{fig:qgan_curves} shows the evolution of various metrics over the course of the QGAN training to learn the latent space of the trained QAE. The adversarial behaviour is observed in the probability curves in Figure \ref{fig:qgan_prob_curves}. 
The training is stopped at the $110th$ iteration where the following two observations are simultaneous met:
\begin{enumerate}
    \item the discriminator identifies both the \textit{real} and the \textit{fake} states as real with probability very close to $1$ (Figure \ref{fig:qgan_prob_curves}).

    \item the cost function $\mathcal{L}_\text{QGAN}$ converges close to $0.5$ (Figure \ref{fig:qgan_loss_curves}) meaning that the discriminator can distinguish between the the \textit{real} and the \textit{fake} states no better than a fair-coin toss.
\end{enumerate}
Both these observable quantities together are useful to decide a stopping criteria for the QGAN training. Further, as a conclusive observation, Figures \ref{fig:entangled_states_qgan_train_110}, and \ref{fig:qgan_vec_overlap} show that the overlap between the \textit{real} and the \textit{fake} states is indeed close to $1.0$ which is in line with the observations in Figures \ref{fig:qgan_prob_curves} and \ref{fig:qgan_loss_curves}. 

The trained generator now has the ability to reproduce the latent representation in Figure \ref{fig:qae_latent_space} by directly generating latent states $\{ \nu_{K} = \eta_{K} \}$ conditioned on the label $K$ using the following optimal parameters:
\begin{equation}
        \overrightarrow{\theta^{*}_{g}} = \big( 1.126\pi, -0.457\pi \big).
\end{equation}
As depicted in Figure \ref{fig:model_generation}, a generated latent state $\nu_{K}$ can be processed using the \textit{trained} decoder (with the optimal parameters $\overrightarrow{\phi^{*}_{D}}$ in Equation \ref{eq:qae_trained_decoder}) to generate 2-qubit entangled states $\xi_{K} = \sigma_{K}$ in Equation \ref{eq:entangled_input_state}.

This example was primarily motivated to showcase the adversarial learning aspect of this framework. The generative aspect of this framework can be appreciated better in the following example. 
\section{Application Example $\mathbf{2}$: \\
Generating Ground State Energy Profiles of Molecular Hamiltonians}

\label{sec:example_2}

This section investigates the utility of the proposed QGAA model to generate the ground state energy profiles of the parameterized molecular Hamiltonians of the Hydrogen molecule ($\mathrm{H}_2$) and Lithium Hydride ($\mathrm{LiH}$). The details of these Hamiltonians are provided in Appendix~\ref{appendix:molecular_hamiltonians}.
A molecular Hamiltonian is dependent upon various parameters such as the coordinates of the constituent nuclei and electrons. Each unique set of parameters corresponds to a specific molecular configuration, which in turn has an associated ground state. 
The fermionic Hamiltonians of both $\mathrm{H}_2$ and $\mathrm{LiH}$ are parameterized by the interatomic distance \( r \) between their constituent atoms. When the fermionic Hamiltonians are mapped to the qubit or the spin versions, the parameter \( r \) is implicitly encoded in the real coefficients $c_{i} \equiv c_{i}(r)$ associated with each Pauli term in the spin Hamiltonian which in general has the following form:

\begin{equation}
    \mathcal{H}(r) = \sum_i c_i P_i
\end{equation}
where each $P_i \in \{I, X, Y, Z\}^{\otimes n}$ is a tensor product of \( n \)-Pauli operators and $c_i \in \mathbb{R}$. 

The objective is to generate the ground state energy landscape $E(r)$ of the ground state of the target molecules as a function of the interatomic distance $r$, using a set of known ground states which serve as the training data for the QGAA model as depicted in Figure \ref{Training:qae}.
We outline the experimental setup and describe the evaluation metrics employed for assessing the different components of the proposed model.
\subsection{Training the QAE for Compression of Molecular Ground States}
\label{sec:h2_qae}

\begin{figure*}[!tbh]
    \centering
    
    \begin{subfigure}{0.49\textwidth}
        \includegraphics[width=0.95\textwidth]{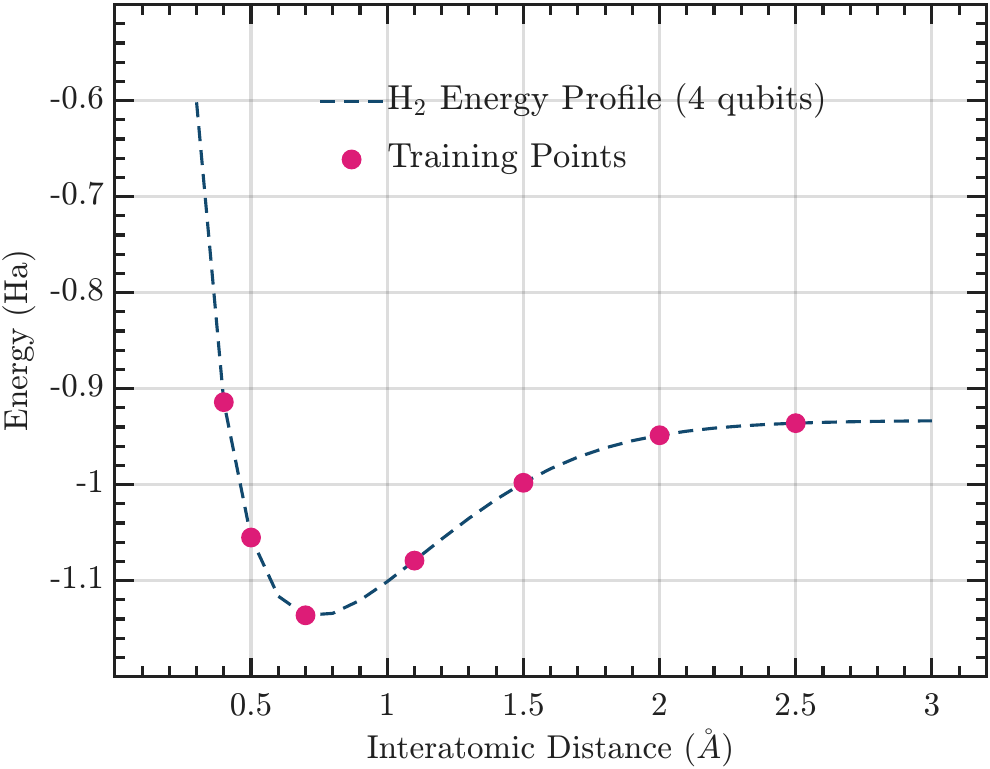}
        \caption{H$_2$}
        \label{H2_energy}
    \end{subfigure}
    \begin{subfigure}{0.49\textwidth}
        \includegraphics[width=0.93\textwidth]{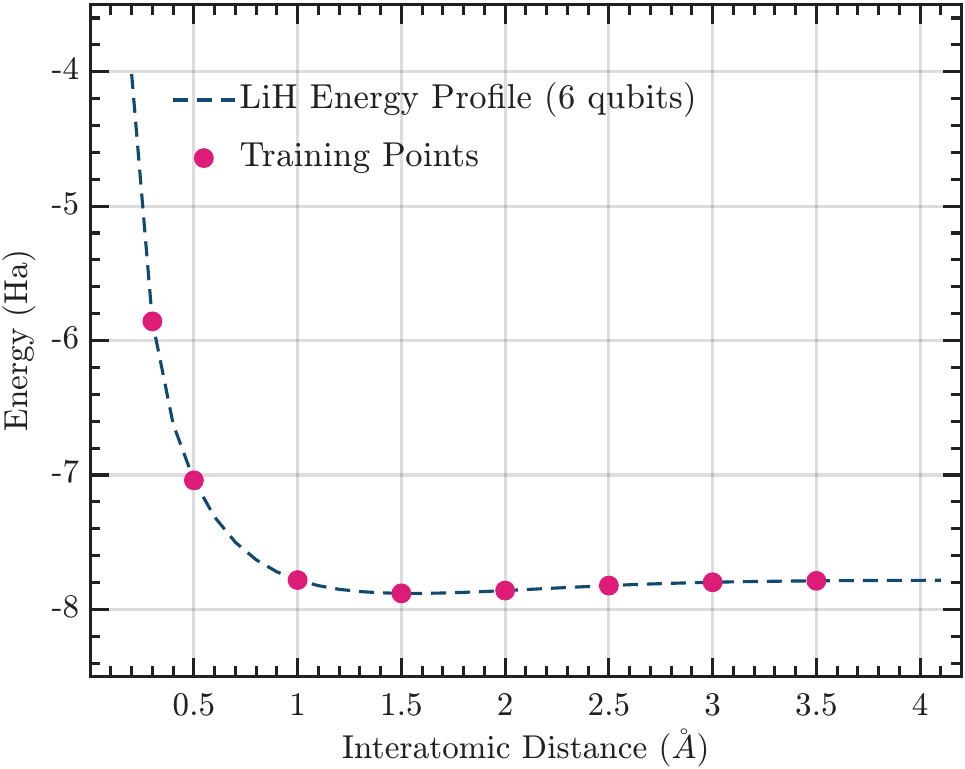}
        \caption{LiH}
    \end{subfigure}
    \caption{Ground state energy profiles across different interatomic distances for the molecules (a) H$_2$ and (b) LiH. The red points indicate the ground states used as training data for the QAE. The QAE is trained to compress molecular ground states of $\mathrm{H}_2$ from $4$ qubits to $1$ qubit, and of LiH from $6$ qubits to $4$ qubit. The performance metrics of the trained QAE are depicted in Table \ref{table:qae_main}.}
    
    \label{Training:qae}
\end{figure*} 

\begin{figure}[!h]
    \centering
    \begin{subfigure}[b]{0.48\columnwidth}
        \centering
        \includegraphics[width=\linewidth]{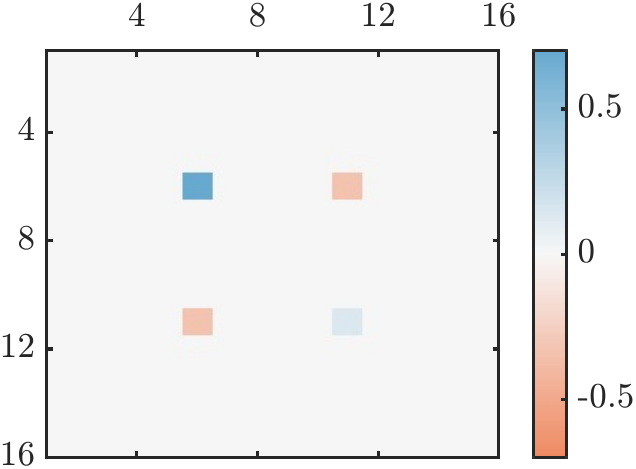}
        \caption{Input State} 
        \label{fig:h2_hm_1.5A}
    \end{subfigure}
    \hfill
    \begin{subfigure}[b]{0.48\columnwidth}
        \centering
        \includegraphics[width=\linewidth]{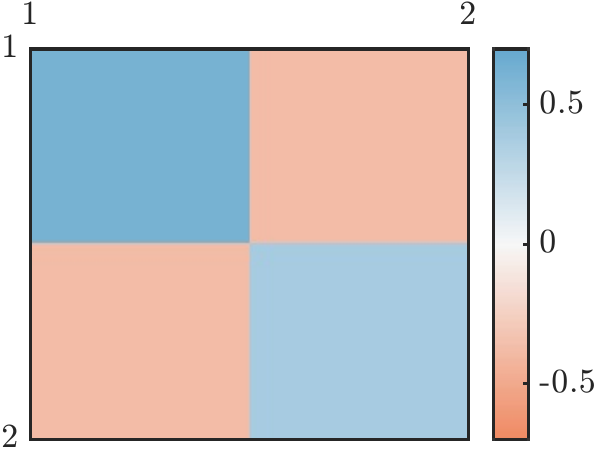}
        \caption{Latent State}
        \label{fig:h2_qae_hm_1.5A}
    \end{subfigure}
    \caption{Heatmaps of the real component of (a) the 4-qubit density matrix of the $\text{H}_{2}$ ground state at $r = 1.5$~\AA{} and (b) its 1-qubit latent representation from a trained QAE.}
    \label{fig:h2_heatmap}
\end{figure}

\begin{figure}[!tbh]
    \centering
    \begin{subfigure}[b]{0.48\columnwidth}
        \centering
        \includegraphics[width=\linewidth]{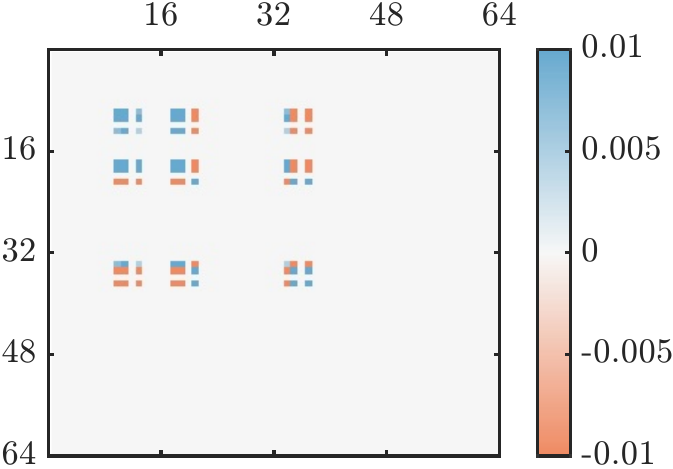}
        \caption{Input State}
        \label{fig:LiH_hm_1.5A}
    \end{subfigure}
    \hfill
    \begin{subfigure}[b]{0.48\columnwidth}
        \centering
        \includegraphics[width=\linewidth]{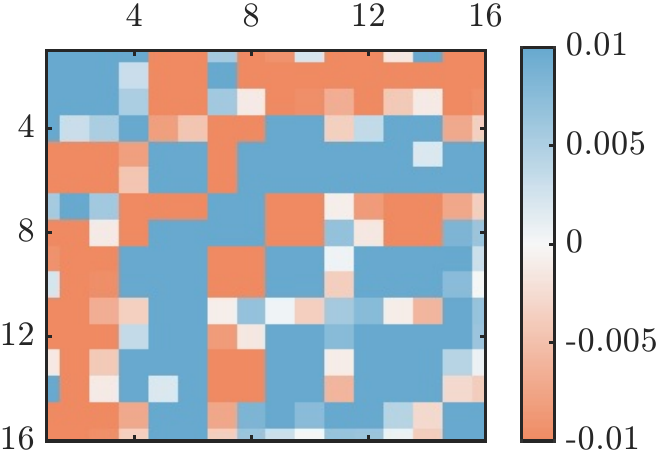}
        \caption{Latent State}
        \label{fig:LiH_qae_hm_1.5A}
    \end{subfigure}
    \caption{Heatmaps of the real component of (a) the 6-qubit density matrix of the $\text{LiH}$ ground state at $r = 1.5$~\AA{} and 
    (b) its 4-qubit latent representation from a trained QAE.}
    \label{fig:LiH_heatmap}
\end{figure}

\begin{figure}[!htb]
    \centering
    \begin{subfigure}[b]{0.45\textwidth}
        \centering
        \includegraphics[width=\textwidth]{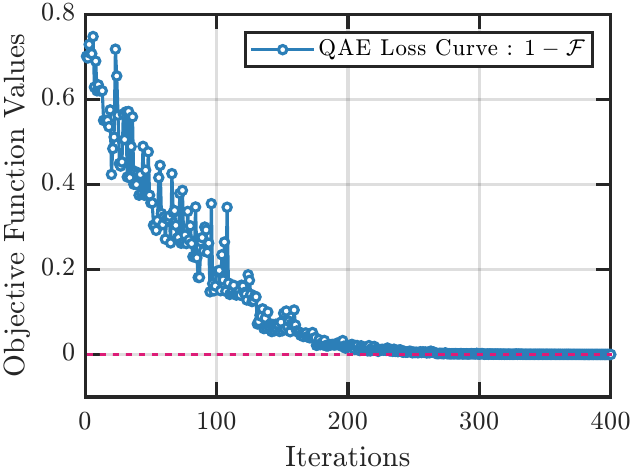}
        \caption{$\text{H}_{2}$}
        \label{fig:H2_QAE_convergence}
    \end{subfigure}
    \hfill
    \begin{subfigure}[b]{0.45\textwidth}
        \centering
        \includegraphics[width=\textwidth]{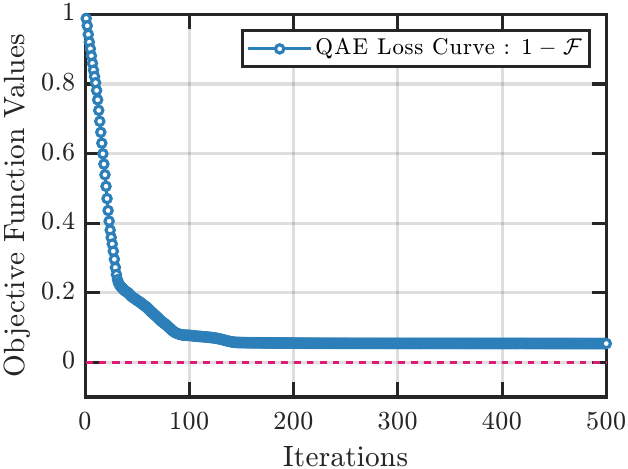} 
        \caption{$\text{LiH}$}
        \label{fig:LiH_QAE_convergence}
    \end{subfigure}
    
    \caption{Convergence of the QAE cost function (Equation~\ref{eq:qae_cost_function}) during training for optimal compression and reconstruction of the molecular ground states of $\text{H}_{2}$ and $\text{LiH}$. The H$_2$ QAE is optimized using the gradient-free COBYLA method, while the LiH QAE is trained using the ADAM  optimizer.}
    \label{fig:molecule_QAE_convergence}
\end{figure}

According to the QGAA formalism, a QAE is first trained to compress the training data into a lower-dimensional latent space. The sparsity in the density matrices, as observed in Figures \ref{fig:h2_hm_1.5A} and \ref{fig:LiH_hm_1.5A} for the $\mathrm{H}_{2}$ and $\mathrm{LiH}$ molecule respectively, captures the possibility of a reduced latent representation of the sparse molecular ground states using QAE \,\cite{romero2017quantum}.

The training dataset $\{ \sigma_{r} \}$ of the QAE comprises $7$ ground states for the case of \textbf{$\mathrm{H}_{2}$} and $8$ ground states for the case of LiH. 
Figure~\ref{Training:qae} depicts the true ground state energy profiles of $\text{H}_{2}$ and $\text{LiH}$. The datapoints corresponding to the training ground states used to train the QAE are also marked in Figure~\ref{Training:qae}.
The label $r$ denotes the interatomic distance for each data point. 
The ground states $\{ \sigma_{r} \}$ using for training the QAE can be prepared on quantum hardware using algorithms such as the Variational Quantum Eigensolver (VQE)~\cite{peruzzo2014variational}.
For our implementation, we selected the compression rates of the molecular ground states as $4$ to $1$ qubit for the case of $\mathrm{H}_{2}$ and $6$ to $4$ qubits for $\mathrm{LiH}$, as detailed in Appendix \ref{appendix:molecular_hamiltonians}. 

The general structure of the ansatz used to implement the encoder and decoder comprises layers of $R_X$ and $R_Y$ rotation gates followed by entangling $CNOT$ gates. The exact form of these ansatzes is depicted in Figures \ref{fig:H2_qae_ansatz} and \ref{fig:LiH_mol_qae_ansatz}, corresponding to \(\mathrm{H}_2\) and \(\mathrm{LiH}\) respectively, in Appendices \ref{appendix:ansatz_H2} and \ref{appendix:ansatz_LiH}.

The parameters of both the encoder and decoder circuits in the QAE are initialized randomly prior to training. To minimize the QAE cost function, the gradient-free 
COBYLA optimizer \cite{powell1994direct} is employed for the case of $\mathrm{H}_{2}$, whereas for the gradient-based ADAM optimizer \cite{kingma2014adam} is employed for the case of $\text{LiH}$.

Table~\ref{table:qae_main} reports the performance metrics of the QAE averaged over the label $r$, namely the fidelity of reconstruction, $\mathcal{F}(\sigma_{r}, \rho_{r})$, and the error in the energy, $|\Delta E(r)| = \big| \text{Tr}\{\mathcal{H}(r)\sigma_{r}\} - \text{Tr}\{\mathcal{H}(r)\rho_{r}\} \big|$. The results depict that the QAE ansatz structure used for the case of $\mathrm{H_2}$ enables the reconstruction of the molecular ground states with high average fidelity, achieving $\mathcal{F}(\sigma_{r}, \rho_{r}) = 0.99$ and an energy reconstruction error of approximately $|\Delta E(r)| \sim 0.1$ mHa. However, in comparison, the performance of the QAE is 
suboptimal for the compression and reconstruction of the $\text{LiH}$ molecular ground states even with an increased number of layers, with an average fidelity of $\mathcal{F}(\sigma_{r}, \rho_{r}) = 0.94 \pm 0.03$ and reconstruction error of $\big \langle |\Delta E (r)| \big \rangle= 0.02 \pm 0.01$. The heatmaps of the compressed latent state representations learned by the trained QAE at a bond length of $r = 1.5$ \AA{} and $r = 2.0$ \AA{} for $\text{H}_{2}$ and $\text{LiH}$ are shown in Figures \ref{fig:h2_qae_hm_1.5A} and \ref{fig:LiH_qae_hm_1.5A}, respectively. Figure~\ref{fig:molecule_QAE_convergence} shows the convergence of the QAE cost function in Equation \ref{eq:qae_cost_function} during training, illustrating the optimization process for achieving optimal compression and reconstruction of the molecular ground states of \(\mathrm{H}_2\) and \(\mathrm{LiH}\).

\begin{table}[!tbh]
    \centering

    \begin{tabular}{c|c|c|c}

        \toprule
 \textbf{Molecule} & Compression Rate & $\big \langle \mathcal{F}(\sigma_{r}, \rho_{r}) \big \rangle$ & {$\big \langle |\Delta E (r)| \big \rangle$ } \\
   &  &  & (Hartree)\\
        \midrule

         ${\mathrm{H_2}}$ &  ${4 \longrightarrow 1}$ {qubits} &  ${0.99}$ & ${\sim 0.0001}$ \\

         $\mathrm{LiH}$ & $6$ $\longrightarrow$ $4$ qubits &  $ 0.94
 \pm 0.03$ & $0.02 \pm 0.01$ \\

        \bottomrule

    \end{tabular}

\caption{Average fidelity error over the ground states $(\mathcal{F})$ 
upon compression and reconstruction using the QAE trained on ground states of \(\mathrm{H_2}\) and \(\mathrm{LiH}\). The average absolute error in the energy of the reconstructed states is also reported.}

\label{table:qae_main}
\end{table}

The encoder of the trained QAE maps each input ground state to a corresponding latent representation. We show that the QAE learns pure-state latent representations of the form $\{ \eta_{r} = \ket{\gamma_{r}}\bra{\gamma_{r}} \}$ for the training states $\{ \sigma_{r} \}$, and enables the reconstruction of the corresponding output states $\{ \rho_{r} \}$ from these latent representations. We note certain caveats in the Quantum Autoencoder formalism, which are discussed in detail in Appendix~\ref{QAE_caveats}. Additional experiments on the LiH Quantum Autoencoder are presented in Appendix~\ref{LiH_QAE_add}. 

As described in Section \ref{sec:adversarial_formalism}, the adversarial training formalism is leveraged in the following section to learn the latent space representation of the trained QAE to generate ground states $\{ \xi_{r} \}$ beyond the training set.

\begin{figure*}[!tbh]
    \centering
    \begin{subfigure}{0.49\textwidth}
        \includegraphics[width=0.95\textwidth]{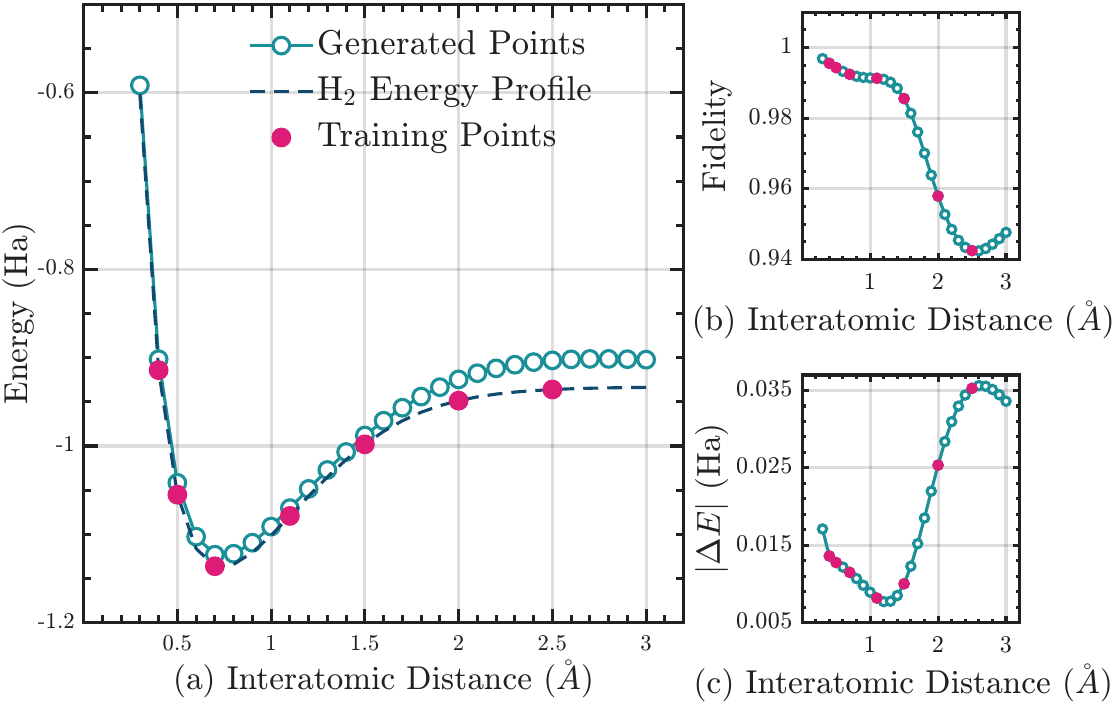}
        \caption{$\mathrm{H_2}$}
        \label{fig:H2_qgan_energy}
    \end{subfigure}
    \begin{subfigure}{0.49\textwidth}
        \includegraphics[width=0.95\textwidth]{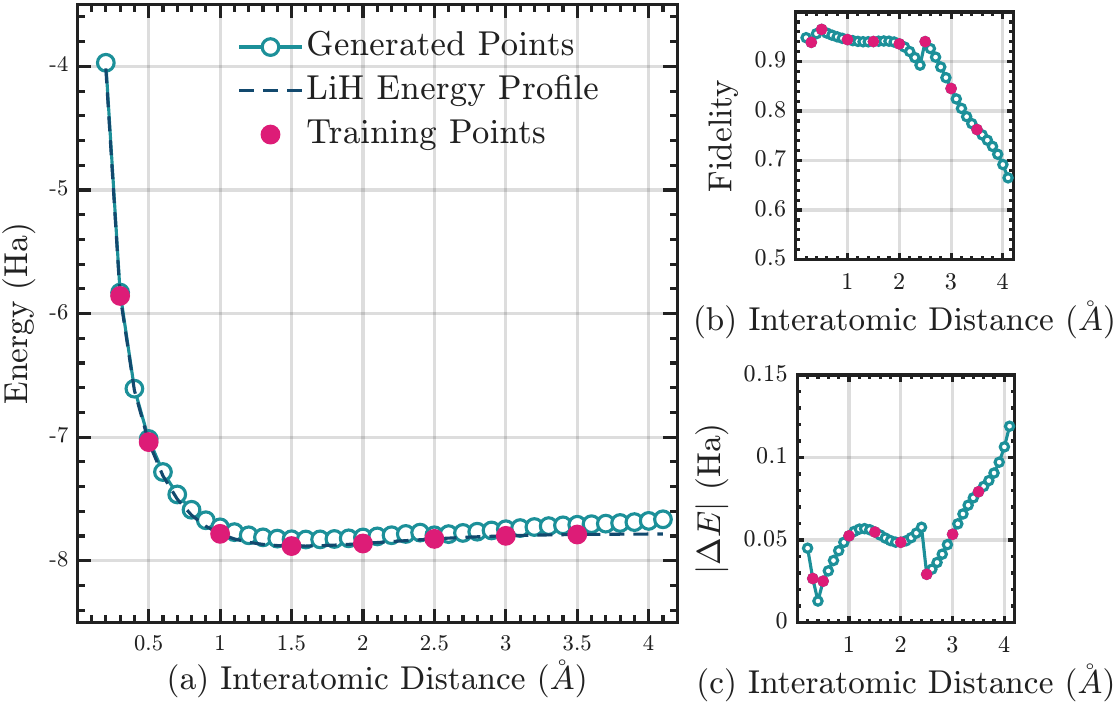}

        \caption{LiH}
        \label{fig:LiH_qgan_energy}
    \end{subfigure}
    \caption{Learned ground state energy profiles across different interatomic distances for the molecules (a) $\mathrm{H_2}$ and (b) $\mathrm{LiH}$. The red points indicate the ground states used as training data for the QGAN. The average fidelity of the generated states is $0.97$ for H$_2$ and $0.88$ for LiH. Panels (b) and (c) in each sub-figure show the fidelity and absolute error for the QGAN-generated states, respectively.}
    \label{fig:qgan_energy_profiles}
\end{figure*}

\subsection{Adversarial Learning of QAE Latent Space for Generation of Molecular Ground States}

\begin{figure}[!thb]
\centering
    \begin{subfigure}{0.45\textwidth}
        \includegraphics[scale=0.75]{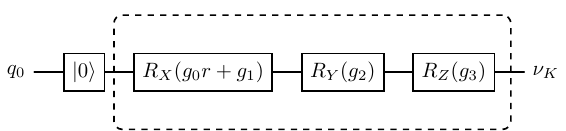}
       \caption{$\text{H}_{2}$}
    \label{fig:mol_gen_ansatz_h2}
    \end{subfigure}
    
    \begin{subfigure}{0.5\textwidth}
        \includegraphics[scale=0.7]{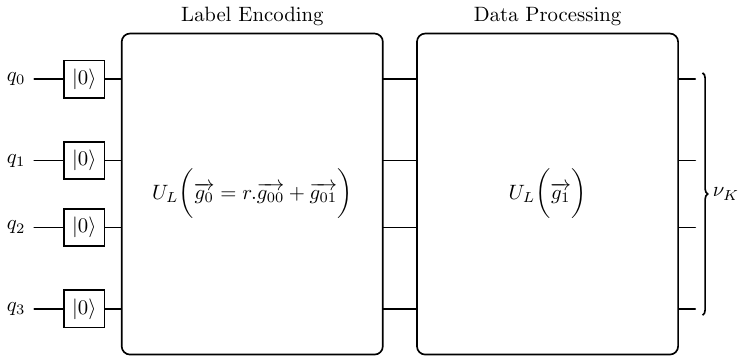}
       \caption{$\text{LiH}$} 
        \label{fig:mol_gen_ansatz_LiH}
    \end{subfigure}

    \caption{Structure of the generator \( U_{g} \big( K, \overrightarrow{\theta_{g}} \big) \) for (a) \(\text{H}_{2}\) and 
    (b) \(\text{LiH}\) - comprising a label encoding block followed by a data processing block \cite{meta-vqe} - with the internal structure of $U_{L}$ detailed in Figure \ref{fig:LiH_gen_ansatz} in Appendix \ref{appendix:ansatz_LiH}.}

    \label{fig:mol_gen_ansatz}
    
\end{figure}

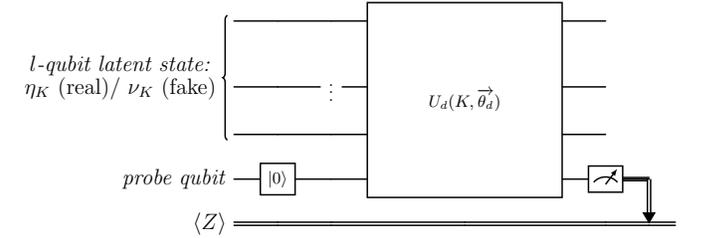
\begin{figure}[!thb]
    \centering
    \scalebox{0.7}{
        \begin{quantikz}[every gate/.append style={minimum height=1.2cm, font=\Large}]
        \lstick[3]{\large \textit{$l$-qubit latent state:} \\ \large $\eta_{K}$ (real)/ $\nu_{K}$ (fake)}&  & &\gate[4][3.7cm]{U_d(K, \overrightarrow{\theta_d})} &\\
         & &\gate[nwires=1,style={fill=white,draw=white,text height=1cm}] {\vdots} & & \\   
         & && &  \\
         \lstick[1]{\large \textit{probe qubit}}&
         \gate{\ket{0}}& & &\meter{}&\cw\arrow[from=4-6, to=5-6, Rightarrow, thick, -{Triangle[scale=0.6]}]
\\
         \lstick[1]{\large $\langle Z \rangle$}&\setwiretype{c}&& & & &
    \end{quantikz}
}

  \caption{Structure of the discriminator \( U_{d} \big( K, \overrightarrow{\theta_{d}} \big) \), which distinguishes between real latent states (\( \eta_{k} \)) and fake latent states (\( \nu_{k} \)). The ansatz structure to implement $U_d(K, \overrightarrow{\theta_d})$ for the case of $\text{H}_{2}$ is the same as depicted in Figure \ref{fig:qgan_disc_ansatz}, and for the case of LiH is detailed in Figure \ref{fig:LiH_mol_qae_ansatz} in Appendix \ref{appendix:ansatz_LiH}.}
    
    \label{fig:disc_structure}
    
\end{figure}

The latent states $\{ \eta_{r} \}$ of the \textit{trained} QAE are the \textit{real data} for the QGAN whereas the states $\{ \nu_{r} \}$ generated by the trainable generator conditioned on the label $r$ are the \textit{fake data}. 
Assuming that the chosen parametrized ansatz for the generator can approximate 
the required solution,
the aim of the adversarial training is to find optimal parameters \(\overrightarrow{\theta^{*}_g}\) such that the \textit{trained} decoder transforms the generated state \(\nu_{r}\)
to a state \(\xi_{r}\) that closely approximates the ground state of the molecular Hamiltonian \(\mathcal{H}(r)\).

The architectures employed for the generator in the cases of $\text{H}_{2}$ and $\text{LiH}$ are shown in Figures\,\ref{fig:mol_gen_ansatz_h2} and\,\ref{fig:mol_gen_ansatz_LiH}, respectively. The architecture used for the discriminator is depicted in Figure\,\ref{fig:disc_structure}. 
The parameters of the generator and the discriminator are randomly initialized and the QGAN is trained in accordance with Algorithm \ref{alg:fully_quantum_adversarial_autoencoder}. The adversarial training employs two separate ADAM optimizers \cite{kingma2014adam} with learning rates of 0.1 and 0.01 for the generator and discriminator, respectively, for the case of $\text{H}_{2}$. Whereas, for the case of $\text{LiH}$, learning rate schedulers are incorporated that reduce the generator’s learning rate by a factor of 0.75 every 100 iterations and the discriminator’s learning rate by the same factor every 250 iterations. This adaptive learning strategy is intended to improve convergence. \\

Table \ref{table:qgan_energy} reports the average fidelity of the generated states, 
$\mathcal{F}(\sigma_{r}, \xi_{r})$, and the error in the energy, 
$|\Delta E (r)| = \big| \text{Tr}\{\mathcal{H}(r)\sigma_{r}\} - \text{Tr}\{\mathcal{H}(r)\xi_{r}\} \big|$, 
evaluated over the interatomic distance $r$. 
The learned ground state energy profile is shown for the $\mathrm{H_2}$ molecule in Fig.~\ref{fig:H2_qgan_energy} and for the $\mathrm{LiH}$ molecule in Fig.~\ref{fig:LiH_qgan_energy}.
The results demonstrate that the trained QGAN closely approximates the true energy landscape, achieving an average state fidelity of $\mathcal{F}(\sigma_r, \xi_r) = 0.97 \pm 0.02$ and an energy reconstruction error of $|\Delta E(r)| = 0.02 \pm 0.01$ Ha for the case of \text{$H_{2}$}. In contrast, the results for LiH are comparatively suboptimal, primarily due to limitations in the trained encoder, as reported in Table \ref{table:qae_main}. These discrepancies highlight the influence of both model architecture and training configuration in the QAE and QGAN frameworks on performance outcomes, especially in the context of scaling to larger molecular systems. We compare the QGAA with a baseline QGAN trained directly on the 4-qubit ground states of H$_2$, with results presented in Appendix~\ref{QGANvsQGAA}, highlighting the utility of using the QAE to achieve more stable and efficient training of the QGAN.

\begin{table}[!tbh]
    \centering

    \begin{tabular}{c|c|c}

        \toprule
 \textbf{Molecule} &  $ \big \langle \mathcal{F}(\sigma_{r}, \xi_{r}) \big \rangle $ & {$\big \langle |\Delta E (r)| \big \rangle$ } \\
   &  &  (Hartree)\\
        \midrule

         ${\mathrm{H_2}}$  &  $ {0.97 \pm 0.02}$ & ${0.02 \pm 0.01}$ \\

         $\text{LiH}$ &  $0.88 \pm 0.09$  & $0.06\pm 0.02$    \\

        \bottomrule

    \end{tabular}

\caption{Average fidelity error (\(\mathcal{F}\)) of the QGAA model, i.e., the QGAN trained on the compressed latent space of the ground states of \(\mathrm{H}_2\) and \(\mathrm{LiH}\). The average absolute error in the energy of the generated states is also reported.}

\label{table:qgan_energy}
\end{table}
\section{Conclusion and Future Outlook}

\label{sec:conclusion}

In this work, we proposed the formalism of the \textit{Quantum Generative Adversarial Autoencoder} (QGAA) for generating quantum states with desired features.

By leveraging the quantum adversarial learning approach, we offer a way to directly access the latent space of a trained QAE, and thus impart the QAE with generative capabilities which it inherently lacks. 
Through the following illustrative examples, we demonstrated that the QGAA can successfully capture and reconstruct the quantum latent space of the QAE:
\begin{enumerate}
    \item Learning the latent space of entangled states compressed by a QAE to analyze the adversarial training behaviour consistent with the theory of QGAN. 

    \item Energy profiling in quantum chemistry: learning the ground state energy landscape of the parameterized Hamiltonians, $\mathrm{H}_2$ and $\mathrm{LiH}$.
\end{enumerate}

The examples implemented in Sections \ref{sec:example_1} and \ref{sec:example_2} are demonstrative of the potential for applying QGAA to generative tasks where quantum state compression can be employed for efficient utilization of resources in QML algorithms.

In our examples, we demonstrate the use of a QGAN to learn the latent representation of the QAE, yielding two key benefits. First, it endows the QAE with generative capabilities by providing direct access to its latent space through the QGAN's generator. Second, the inherent ability of the QAE to compress quantum data into a lower dimensional latent space enables a more resource efficient implementation of the QGAN. Thus, the QGAA has a bidirectional advantage of enhancing the QAE with generative functionality while simultaneously reducing the resource costs for a QGAN. 
 
Notably, the training procedure only requires access to quantum states, and the resource requirements for training a QGAA on a compressed subspace are significantly lesser than that for training a QGAN on the original larger space of the quantum states. 

Although the theory of quantum adversarial learning guarantees the existence of a unique Nash equilibrium, we encountered practical challenges in steering the model toward this optimal solution during training. Nevertheless, our results show that even with standard optimization techniques, it is possible to reasonably learn the latent space representation and generate quantum states that closely approximate the target states. In this context, various QGAN variants have been proposed in the literature, which could be integrated into our framework to stabilize training dynamics and improve optimization efficiency \cite{wqgan, quantum_w1_distance, local_quantum_w1_distance}. 

Even upon learning the latent representation approximately and not exactly, QGAA has utility in quantum machine learning applications such as warm-starting another quantum algorithm like the VQE with a better guess of the initial state \cite{meta-vqe}. The generated ground states may serve as useful initial points that could potentially lead to more efficient convergence to the optimal solution \cite{meta-vqe}.

We also highlighted the nuances and the assumptions in the QGAA framework and are explained in Appendices \ref{QAE_caveats} and \ref{appendix:qgan_caveats}. The framework of QGAA proposed in this work thus paves a new direction for exploring the efficient execution of quantum generative tasks.  

\section{Open questions and challenges}

As a QML model, the QGAA 
has several open questions and avenues for future investigation. The following are key aspects to be explored:

\begin{enumerate}

    \item Scalability and Training: The implementations explored in this study were limited to small-scale quantum systems with the largest being $6$ qubits for the $\text{LiH}$ molecule. 
    Scaling the QGAA to larger-qubit systems is challenging both for the QAE and the QGAN. Overcoming the issues posed by vanishing-gradients \cite{ragone2024lie, bhowmick2025enhancing} is critical for resource-efficient and stable training of QML models of increasing system sizes. Particularly for the QAE component, the SWAP test (employed to estimate the overlap of the reconstructed state) is computationally expensive to evaluate with scaling system size. The utility of methods such as approximate fidelity estimation using shadow tomography techniques can be explored to address the issues associated with fidelity-based loss functions \cite{huang2020predicting}. 
    
    \item Hardware implementation: While our experiments were conducted on simulators, deploying the proposed model on real quantum hardware introduces additional considerations such as gate noise, decoherence, and limited qubit connectivity. These hardware-induced imperfections can significantly affect the performance of both the QAE and QGAN components. Fine-tuning the model for useful performance on quantum devices requires the incorporation of noise-resilient circuit designs, error mitigation techniques, and hardware-aware optimization strategies to ensure reliable implementation under realistic conditions.
    
\end{enumerate}

\section{Code Availability}

The code developed to obtain the results in this paper will be made available at a later date. 

\begin{acknowledgments}
The authors acknowledge the role of Fujitsu Research in supporting this project. We are also grateful to our colleagues - Ruchira Bhat, Rahul Bhowmick, Harsh Wadhwa, Aritra Sarkar, and Quoc Hoan Tran for their insights and valuable feedback that helped develop this work. 
The authors extend their immense gratitude to Yasuhiro Endo,
Hirotaka Oshima and Shintaro Sato, as well as the entire Robust Quantum Computing Department at Fujitsu
Limited for their strategic and technical support.
A preliminary version of this paper was presented
at the Fujitsu-IISc Quantum Workshop in Bangalore, India, and a poster presentation is scheduled for TQC 2025, to be held from September 15–19.
\end{acknowledgments}

\bibliographystyle{plainnat}
\bibliography{references}
\onecolumngrid
\appendix

\newpage

\section{Autoencoders}

\label{sec:ae_appendix}

\begin{figure}[!tbh]
        \centering
        \includegraphics[width=0.8\linewidth]{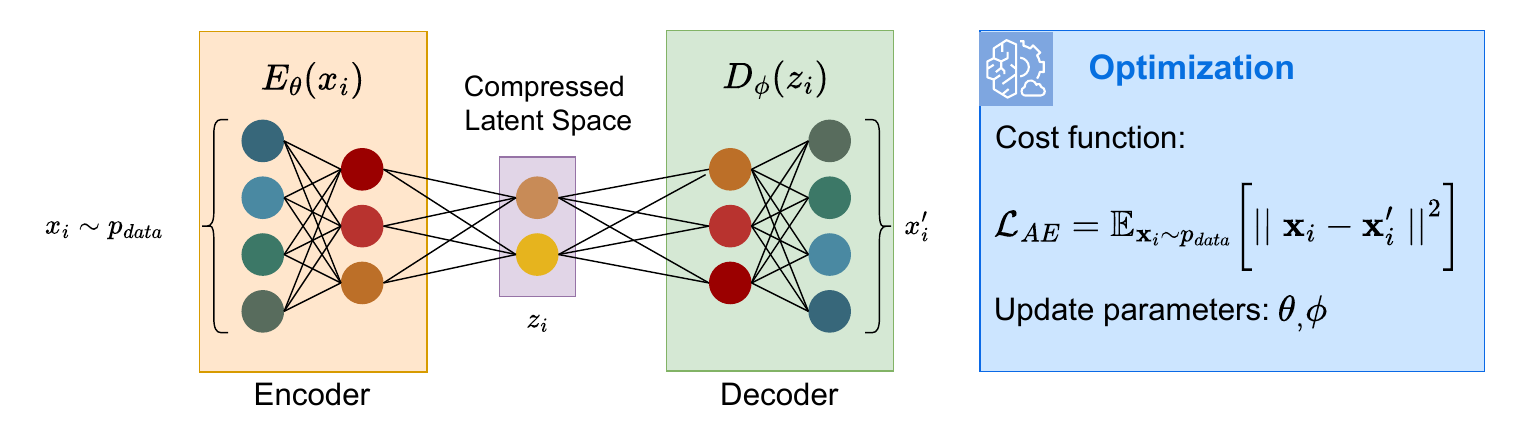}
        \caption{Schematic of the Autoencoder (AE). The training set is denoted as $\mathbf{X}_{\text{train}} = \{ \mathbf{x}_{i}\}_{i=1}^{n}$, where each input $\mathbf{x}_{i} \in \mathbb{R}^{d}$ is a $d$-dimensional vector sampled from the data distribution $p_{\text{data}}$. The encoder $E_{\bm{\theta}}: \mathbb{R}^{d} \mapsto \mathbb{R}^{m}$, with parameters $\bm{\theta}$, compresses each input into a latent vector $\mathbf{z}_{i} \in \mathbb{R}^{m}$ with $m < d$. The decoder $D_{\bm{\phi}}: \mathbb{R}^{m} \mapsto \mathbb{R}^{d}$, with parameters $\bm{\phi}$, maps $\mathbf{z}_{i}$ back to the reconstructed output $\mathbf{x}_{i}^{\prime} \in \mathbb{R}^{d}$. Training minimizes the reconstruction loss $\mathcal{L}_{\text{AE}} = \mathbb{E}_{\mathbf{x}_{i} \sim p_{\text{data}}}[ \lVert \mathbf{x}_{i} - \mathbf{x}^{\prime}_{i} \rVert^{2} ]$, ensuring that $\mathbf{x}_{i}^{\prime}$ remains close to $\mathbf{x}_{i}$. While effective for compression, the AE does not possess generative capability.}

        \label{fig:AE_schematic}
\end{figure}

An autoencoder (AE) \cite{ae_hinton_learning_internal_representations} is a machine learning architecture designed to compress input data into a latent representation, as depicted schematically in Figure~\ref{fig:AE_schematic}. AE has been applied in various domains such as dimensionality reduction, classification, anomaly detection, and denoising \cite{ae_denoising}. Consider a training set $\mathbf{X}_{\text{train}} = \{ \mathbf{x}_{i}\}_{i=1}^{n}$ sampled from some data distribution $p_{\text{data}}$, where $n$ is the number of training points and each $\mathbf{x}_i \in \mathbb{R}^{d}$ is a $d$-dimensional input vector. The architecture consists of two classical neural networks. The encoder $E_{\bm{\theta}}: \mathbb{R}^{d} \mapsto \mathbb{R}^{m}$ with trainable parameters $\bm{\theta}$ maps an input vector $\mathbf{x}_{i}$ to a compressed latent representation $\mathbf{z}_{i} = E_{\bm{\theta}}(\mathbf{x}_{i}) \in \mathbb{R}^{m}$, where $m < d$ ensures compression. The decoder $D_{\bm{\phi}}: \mathbb{R}^{m} \mapsto \mathbb{R}^{d}$ with parameters $\bm{\phi}$ maps this latent vector back to the input space, reconstructing the original data point as $\mathbf{x}^{\prime}_{i} = D_{\bm{\phi}}(\mathbf{z}_{i})$. Training is carried out by minimizing the reconstruction loss, defined as the average squared error between the input and its reconstruction,
\begin{equation}
    \underset{\bm{\theta}, \bm{\phi}}{\min} \; \mathcal{L}_{\text{AE}} := \mathbb{E}_{\mathbf{x}_{i} \sim p_{\text{data}}} \big[ \lVert \mathbf{x}_{i} - \mathbf{x}^{\prime}_{i} \rVert^{2} \big],
    \label{eq:ae_cost_02}
\end{equation}
which ensures that the reconstructed outputs $\mathbf{x}_{i}^{\prime} = D_{\bm{\phi}}(E_{\bm{\theta}}(\mathbf{x}_{i}))$ remain close to the original inputs $\mathbf{x}_{i}$. The AE is a deterministic model that effectively compresses data into a lower-dimensional latent space but lacks generative capability. The Variational Autoencoder (VAE) \cite{kingma2013auto, vae_book_kingma}, explained in the Appendix \ref{vae_appendix}, extends this framework by enabling both compression and the generation of new samples not present in the training set $\mathbf{X}_{\text{train}}$.


\section{Metric Used to evaluate the QAE Cost function}\label{metrics}

The SWAP test is a standard quantum subroutine used to estimate the overlap between two quantum states 
\(\sigma_{K}, \rho_{K}\). 
Given two states, the quantity of interest is the Hilbert–Schmidt inner product
\begin{equation}
    \text{SWAP}(\sigma_{K}, \rho_{K}) := \mathrm{Tr} \big( \sigma_{K} \, \rho_{K} \big).
\end{equation}

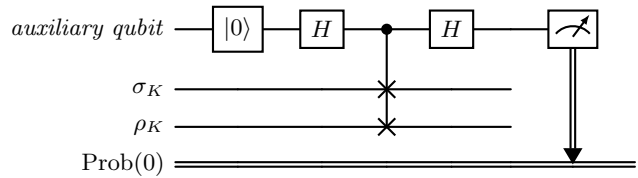
\begin{figure}[!thb]
    \centering
    \scalebox{1}{
        \begin{quantikz}
        \lstick[1]{\textit{auxiliary qubit}} & \gate{\ket{0}} & \gate{H} & \ctrl{1}& \gate{H} & &\meter{}\arrow[from=1-7, to=4-7, Rightarrow, thick, -{Triangle[scale=0.6]}]\\
         \lstick[1]{$\sigma_{K}$} &  &  & \swap{1}& & \\
          \lstick[1]{$\rho_{K}$} &  &  &\swap{-1} &  &\\
          \lstick[1]{\text{Prob($0$)}}&\setwiretype{c}&& & & & &
    \end{quantikz}
}
  \caption{SWAP Test for evaluating overlap between the quantum states $\sigma_{K}$ and $\rho_{K}$. \text{Prob($0$)} is the probability of measuring the auxiliary qubit in the state $\ket{0}$.}
    \label{fig:swap_test_schematic}
\end{figure}
The protocol, illustrated in Figure~\ref{fig:swap_test_schematic}, introduces an auxiliary qubit, initialized in the state $\ket{0}$, which controls a SWAP operation between the two registers containing $\sigma_{K}$ and $\rho_{K}$. 
After applying Hadamard gates before and after the controlled-SWAP, the auxiliary qubit is measured in the computational basis. 
The probability of observing outcome $\ket{0}$ is given by,
\begin{equation}
    \mathrm{Prob}(0) = \tfrac{1}{2}\big(1 + \mathrm{Tr}(\sigma_{K}\rho_{K})\big).
\end{equation}
Rearranging, one obtains
\begin{equation}
    \text{SWAP}(\sigma_{K}, \rho_{K}) = \mathrm{Tr}(\sigma_{K} \rho_{K}) 
    = 2 \, \mathrm{Prob}(0) - 1.
\end{equation}

Thus, the SWAP test provides an operational way to estimate the overlap between two density operators by repeated sampling of the ancilla measurement outcome. \\


\textbf{Fidelity:} For two general (possibly mixed) states $\rho_k$ and $\sigma_k$,  the \textit{Uhlmann–Jozsa fidelity} \cite{Jozsa01121994, liang2019quantum} is
\begin{equation}
    \mathcal{F}(\sigma_k, \rho_k) = \left( \mathrm{Tr}\, \sqrt{ \sqrt{\sigma_k}\,\rho_k\,\sqrt{\sigma_k} } \right)^2,
\end{equation}
which reduces to $|\braket{\psi_k|\phi_k}|^2$ for two pure states and to $\bra{\psi_k}\rho_k\ket{\psi_k}$ when $\sigma_k=\ket{\psi_k}\bra{\psi_k}$ is pure.




\noindent
\textbf{Key properties:}
\begin{itemize}[noitemsep, nolistsep]
    \item \( 0 \leq \mathcal{F} \leq 1 \)
    \item \( \mathcal{F} = 1 \) if and only if the states are identical
    \item Fidelity is symmetric: \( \mathcal{F}(\rho_k, \sigma_k) = \mathcal{F}(\sigma_k, \rho_k) \)
    \item Fidelity is invariant under unitary transformation $U$ such that  \( \mathcal{F}(\rho_k, \sigma_k) = \mathcal{F}(U\rho_k U^{\dagger}, U\sigma_k U^{\dagger})\). 
\end{itemize}

\section{Variational Autoencoders}
\label{vae_appendix}
\begin{figure}[!tbh]
        \centering
         \includegraphics[width=0.9\linewidth]{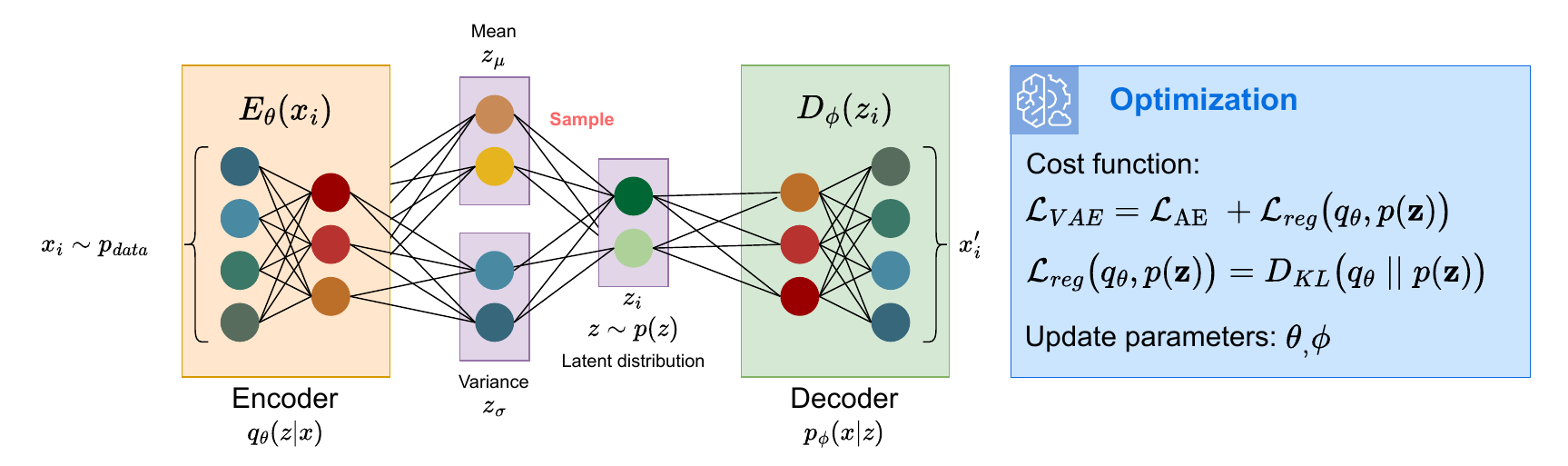}
        \caption{Schematic of the Variational Autoencoder (VAE). An input datapoint $\mathbf{x}_{i} \in \mathbb{R}^{d}$ from the training set is mapped by the encoder $E_{\bm{\theta}}$ into the parameters of a latent probability distribution $q_{\bm{\theta}}(\mathbf{z}|\mathbf{x}_{i})$ over latent variables $\mathbf{z}_{i} \in \mathbb{R}^{\ell}$, where $\ell < d$. Unlike a deterministic Autoencoder, which encodes each input into a single latent vector, the VAE samples $\mathbf{z}_{i}$ from this distribution. The latent distribution is regularized to match a chosen prior $p(\mathbf{z})$, typically the standard normal $\mathcal{N}(\mathbf{0}, \mathbb{I})$, ensuring that the latent space is continuous and well-structured. The decoder $D_{\bm{\phi}}$ then maps a sampled latent vector $\mathbf{z}_{i}$ back to the data space, producing a reconstruction $\mathbf{x}_{i}^{\prime} = D_{\bm{\phi}}(\mathbf{z}_{i})$. This probabilistic formulation enables the VAE to both reconstruct training data and generate novel data points $\mathbf{x}_{i^{\prime}}$ by sampling $\mathbf{z}_{i^{\prime}} \sim p(\mathbf{z})$.
}

        \label{fig:VAE_schematic}
\end{figure}

Unlike an Autoencoder (AE), which is a deterministic model that maps input data into a discrete latent representation, a Variational Autoencoder (VAE) \cite{kingma2013auto, vae_book_kingma} is a probabilistic model, as shown schematically in Figure~\ref{fig:VAE_schematic}. Instead of encoding the latent variables $\mathbf{z}$ of training data as fixed points, the VAE encodes them as a probability distribution. Specifically, the encoder outputs an approximate posterior distribution $q_{\bm{\theta}}(\mathbf{z}|\mathbf{x})$, rather than a single latent vector, while the prior distribution over latent variables is denoted as $p(\mathbf{z})$. This ensures that the latent space has a continuous probabilistic structure. Once trained, the model can sample $\mathbf{z}$ from $p(\mathbf{z})$ and generate new datapoints through the decoder. The architecture of a VAE resembles that of an AE, consisting of an encoder $E_{\bm{\theta}}$ that maps an input $\mathbf{x}_{i} \in \mathbf{X}_{\text{train}}$ to a latent distribution $q_{\bm{\theta}}(\mathbf{z}|\mathbf{x}_{i})$, and a decoder $D_{\bm{\phi}}$ that reconstructs or generates data as $\mathbf{x}_{i}^{\prime} = D_{\bm{\phi}}(\mathbf{z}_{i})$, where $\mathbf{z}_{i}$ is sampled from $q_{\bm{\theta}}(\mathbf{z}|\mathbf{x}_{i})$. Training involves not only minimizing the reconstruction loss $\mathcal{L}_{\text{AE}}$ (Equation~\ref{eq:ae_cost_02}), but also regularizing the latent distribution $q_{\bm{\theta}}(\mathbf{z}|\mathbf{x})$ so that it approximates a known prior $p(\mathbf{z})$, typically chosen as the standard normal distribution $\mathcal{N}(\mathbf{0}, \mathbb{I})$. The resulting loss function is  
\begin{equation}
    \underset{\bm{\theta}, \bm{\phi}}{\min} \; \mathcal{L}_{\text{VAE}} := \mathcal{L}_{\text{AE}} + \mathcal{L}_{\text{reg}}\big(q_{\bm{\theta}}(\mathbf{z}|\mathbf{x}), p(\mathbf{z})\big),
\end{equation}
where the regularization term $\mathcal{L}_{\text{reg}}$ measures the divergence between the approximate posterior $q_{\bm{\theta}}(\mathbf{z}|\mathbf{x})$ and the prior $p(\mathbf{z})$, defined using the Kullback–Leibler (KL) divergence as  
\begin{equation}
\label{eq:vae_cost_02}
    \mathcal{L}_{\text{reg}}\big(q_{\bm{\theta}}(\mathbf{z}|\mathbf{x}), p(\mathbf{z})\big) 
    = D_{\mathrm{KL}}\big(q_{\bm{\theta}}(\mathbf{z}|\mathbf{x}) \,\Vert\, p(\mathbf{z})\big).
\end{equation}

At convergence, $q_{\bm{\theta}}(\mathbf{z}|\mathbf{x})$ aligns with $p(\mathbf{z})$, ensuring that the latent space is continuous and structured. As a result, new latent samples $\mathbf{z}_{i^{\prime}} \sim p(\mathbf{z})$, which do not correspond to any encoded training datapoint, can be passed through the trained decoder to generate novel datapoints $\mathbf{x}_{i^{\prime}} = D_{\bm{\phi}}(\mathbf{z}_{i^{\prime}})$. Thus, unlike deterministic AE which is limited to reconstructing training data, VAE possesses true generative capability by directly sampling from the latent space to create new data beyond the training set.

\section{Generative Adversarial Networks}

\label{sec:gan_appendix}

\begin{figure}[!tbh]
        \centering
         \includegraphics[width=\linewidth]{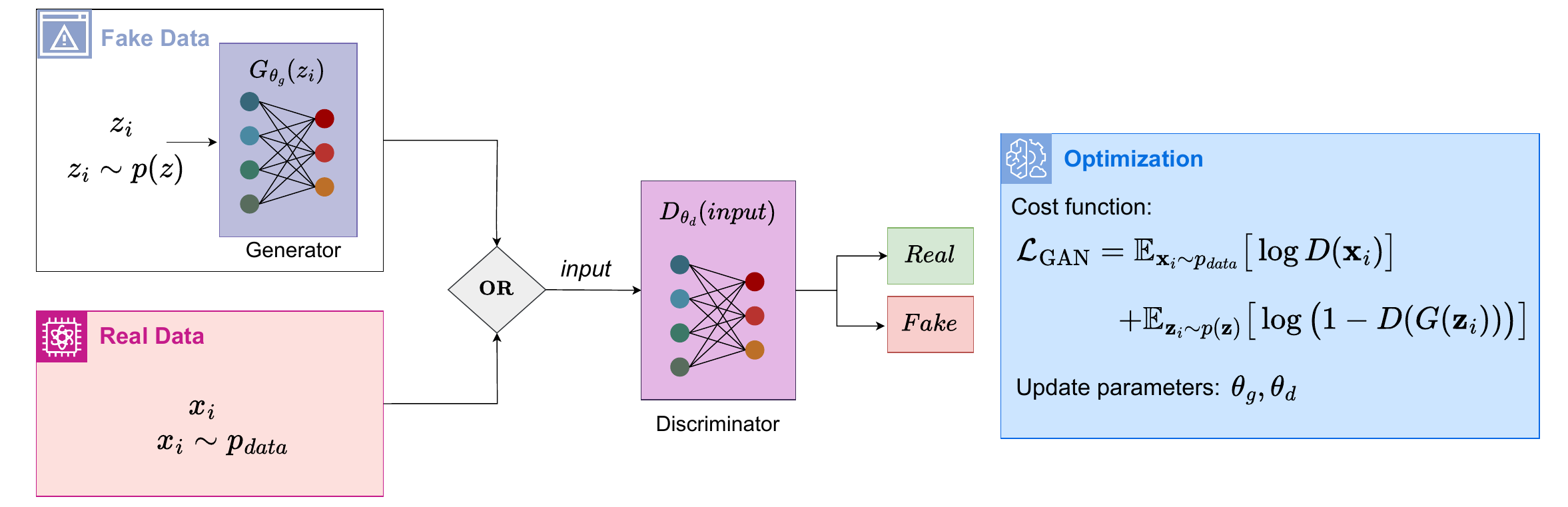}
        \caption{Schematic of the Generative Adversarial Network (GAN). The min-max optimization of $\mathcal{L}_\text{GAN}$ by the generator and the discriminator characterizes the adversarial training. At the Nash equilibrium of the training, the generator generates \textit{fake} data samples $G(\mathbf{z}_{i})$ that the discriminator cannot distinguish from the \textit{real} data samples $\mathbf{x}_{i} \sim p_{data} (\mathbf{x})$.}

        \label{fig:GAN_schematic}
        
\end{figure}

A Generative Adversarial Network (GAN) \cite{gan_original} is a generative framework where two neural networks — a generator $G$ and a discriminator $D$ with parameters $\bm{\theta_{g}}$ and $\bm{\theta_{d}}$ respectively — are trained in an adversarial manner, as shown schematically in Figure~\ref{fig:GAN_schematic}. Conditioned on some prior entropy source $p(\mathbf{z})$, the objective of G is to produce samples of synthetic data $G \big( \mathbf{z}_{i} \sim p(\mathbf{z}) \big)$ that is indistinguishable from samples of the training data $\mathbf{x}_{i} \sim p_{data} (\mathbf{x})$. 
On the other hand, the objective of D is to correctly identify with high probability whether the input data supplied to it is from the real distribution $p_{data}$ or from the prior $p(\mathbf{z})$ to the generated/fake distribution. In other words, $G$ generates synthetic data, and D outputs has a probabilistic binary outcome.

The way the training is setup is such that $G$ and $D$ compete against each other in an adversarial game until they reach the Nash equilibrium or the fixed-point. In general, $D(.)$ outputs a score corresponding to the input data and the maximizing objective of $D(.)$ is to assign a higher score to real data samples $\mathbf{x}_{i} \sim p_{data}$ and a lower score to generated/fake data samples $G(\mathbf{z}_{i})$ where $\mathbf{z}_{i} \sim p(\mathbf{z})$. The minimizing objective of $G(.)$ is to generate samples $G(\mathbf{z}_{i})$ such that $D(.)$ assigns them a higher score. This min-max optimization of the two competing objectives is captured by the cost function $\mathcal{L}_{\text{GAN}}(G,D)$ in Equation \ref{eq:gan_cost}. 
\begin{equation} 
\label{eq:gan_cost}
    \begin{split}
        \underset{\bm{\theta_{g}}}{min} \; \underset{\bm{\theta_{d}}}{max} \; \mathcal{L}_{\text{GAN}} :=
        \mathbb{E}_{\mathbf{x}_{i} \sim p_{data}(\mathbf{x})} \big[ \log{D(\mathbf{x}_{i})} \big] \\
        + \;  \mathbb{E}_{\mathbf{z}_{i} \sim p(\mathbf{z})} \big[ \log \big( 1 - D(G(\mathbf{z}_{i}) \big) \big]
    \end{split}
\end{equation}
The Nash equilibrium of this adversarial game is reached when the features of the generated data are close to that of the real data, at which point the discriminator can no longer discriminate between the real and generated/fake data \cite{gan_original}. \\


\section{Caveats in the Quantum Generative Adversarial Networks formalism}
\label{appendix:qgan_caveats}
The assumptions mentioned in Section \ref{sec:QGAN} with regards to the generator, $G$, and the discriminator, $D$, are detailed further as follows \,:
\begin{enumerate}
    \item \textit{First, it is assumed that both the quantum information processors $G$ and $D$ have enough} \textit{capacity} \cite{eqgan} \textit{or} \textit{expressibility} \cite{expressibility} \textit{such that they can approximate any arbitrary function or transformation using their parameters $\overrightarrow{\theta_{g}}$ and $\overrightarrow{\theta_{d}}$ respectively.} \\ \\
    In practice, this requirement is assumed to be fulfilled if the structure of $G$ and $D$ is a deep quantum feature map that
    is \textit{expressive enough} for $G$ and $D$ to achieve their respective optimal strategies. 
    These feature maps are often chosen heuristically and typically comprise blocks of rotation gates followed by entangling gates. 
    It can be reasoned that ansatzes with such a structure are \textit{highly versatile} \cite{apoorva_density_matrix} and hence are able to create a large variety of feature maps based on different parameterizations, some of which can approximate the desired transformation.
    This ansatz structure can be easily generalized to higher qubits and can be made \textit{deep} or \textit{shallow} based of the number of repeating blocks composed. However, the effectiveness of the chosen ansatz remains dependent on the problem and the physical realization of the gates.

    \item \textit{Second, it is assumed that an efficient optimization scheme exists that can drive the algorithm to converge to the unique Nash equilibrium.} \\
    

Although the theory of Helstrom measurement helps in setting up the quantum adversarial game, it does not give any guarantees on the convergence to the Nash equilibrium.
Analytically, at a given  $(i)th$ iteration of the game the \textit{optimal strategy of} $D$ is to optimize $\overrightarrow{\theta_{d}}$ such that operator $\hat{T}^{(i)}$ projects onto the positive eigenspace of $\big( \sigma - \rho^{(i-1)} \big)$ thus maximizing $\mathcal{L}_{\text{QGAN}}$ \cite{wilde2013quantum}. In practice, this is difficult to achieve since the state $\sigma$ is not known apriori. That is the exact reason why the adversarial protocol is being executed so that the statistics of $\sigma$ could be learned. However, let's assume that $D$ can somehow achieve this particular analytical solution in practice. Then in the next $(i+1)th$ iteration, the analytical \textit{optimal strategy of} $G$ is to optimize $\overrightarrow{\theta_{g}}$ such that the form of the density matrix of the state $\rho^{(i+1)}$ is a pure state that projects onto the largest eigenvalue of $\hat{T}^{(i)}$ thus minimizing $\mathcal{L}_{\text{QGAN}}$. This, however, is not useful for learning mixed-states in general \cite{qgan_analytical}.  \\ \\
Such \textit{exact optimization} at every iteration can potentially lead to \textit{mode collapse} or oscillations between some quantum states \cite{eqgan} and \textit{limit cycles} \cite{qgan_analytical}. 
One approach towards addressing \textit{mode collapse} is by using the square of $\mathcal{L}_{\text{QGAN}}$ as the cost function instead \cite{hamiltonian_qgan}. 
Other efforts include the development of quantum extensions of the classical Wasserstein GAN \cite{wqgan, quantum_w1_distance, local_quantum_w1_distance}. \\ \\
Further, deep feature maps are argued to be powerful enough to approximate arbitrary transformations, and besides the QGAN-specific challenges discussed so far, vanishing gradients or barren plateaus is another significant challenge in training such deep feature maps
\cite{barren_plateau_qgan_zoufal}. Although a unique fixed-point exists in the quantum adversarial game \cite{qgan_01}, efficient optimization techniques to converge to the fixed-point is a challenging and open problem. Studies training fully-quantum GANs to generate quantum states are are limited to $1 -3$ qubits both on simulators \cite{qgan_02, hamiltonian_qgan, eqgan, qgan_analytical} and real devices \cite{qgan_superconducting_bloch, qgan_superconducting_multiple}.

\end{enumerate}

\section{Adversarial Autoencoder}

\label{appendix:adversarial_autoencoder}

\begin{figure}[!tbh]
    \centering
    \includegraphics[width=0.9\linewidth]{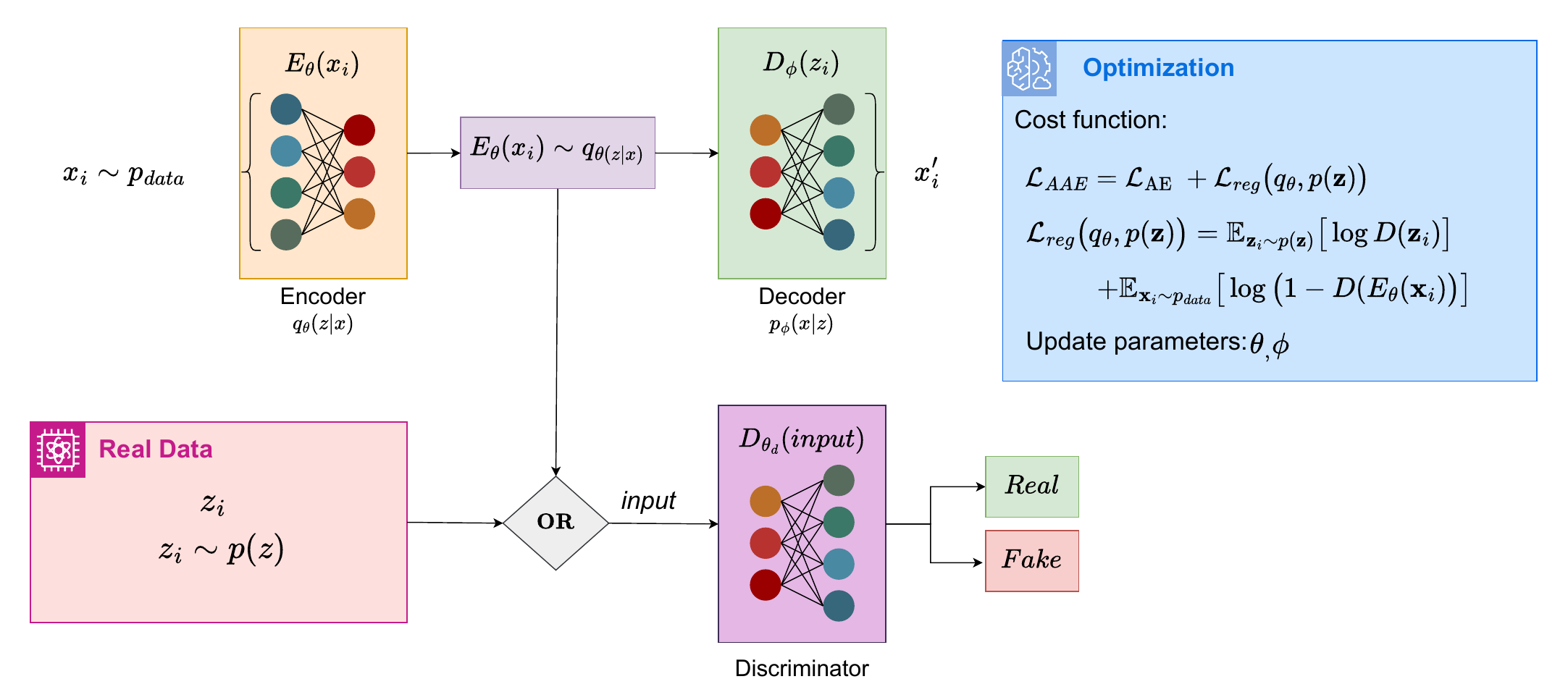}
    \caption{Schematic of the Adversarial Autoencoder (AAE). Unlike the VAE in Figure \ref{fig:VAE_schematic}, the regularization of the latent space in this case is performed using adversarial training instead of evaluating the KL-divergence.}
    \label{fig:AAE_schematic}

\end{figure}


The original framework of the VAE requires that the form of $p(\mathbf{z})$ be known, as mentioned previously in Appendix \ref{vae_appendix}. 
 $p(\mathbf{z})$ can be quantified and described by measurable quantities of the distribution like mean and variance in the case of a normal distribution. In practice, these quantities are evaluated during training (Figure \ref{fig:VAE_schematic}) to regularize the latent distribution $q_{\theta}(\mathbf{z}|\mathbf{x})$ to the known distribution $p(\mathbf{z})$. However, the Adversarial Autoencoder (AAE) \cite{adversarial_autoencoder} was developed to show that the adversarial formalism of a GAN can also be employed to regularize the latent distribution $q_{\bm{\theta}}(\mathbf{z}|\mathbf{x})$ with the target distribution $p(\mathbf{z})$, as shown schematically in Figure~\ref{fig:AAE_schematic}. Here the encoder $E_{\theta}$ acts as the generator that generates a sample $E_{\theta}(\mathbf{x}_{i}) \sim q_{\theta}(\mathbf{z}|\mathbf{x})$ of the latent space given input $\mathbf{x}_{i}$. Another neural network $D$ is used as a discriminator to distinguish between samples from $p({\mathbf{z}})$ (real) and $q_{\theta}(\mathbf{z}|\mathbf{x})$ (fake).
The advantage of using a GAN for VAE latent space regularization is that the complete form of the target distribution $p(\mathbf{z})$ is not required to be known. 
Access to $p(\mathbf{z})$ with the ability to sample from it is enough to regularize $q_{\theta}(\mathbf{z}|\mathbf{x})$ using the adversarial protocol.
Therefore $\mathcal{L}_{reg} \big( q_{\theta}(\mathbf{z}|\mathbf{x}), p(\mathbf{z}) \big)$ (Equation \ref{eq:vae_cost_02}) takes the form represented in Equation \ref{eq:gan_cost} as depicted in Equation \ref{eq:adversarial_ae_cost}:
\begin{equation} 
\label{eq:adversarial_ae_cost}
    \begin{split}
        \underset{\bm{E_{\theta}}}{min} \; \underset{\bm{D}}{max} \; \mathcal{L}_{reg} \big( q_{\theta}(\mathbf{z}|\mathbf{x}), p(\mathbf{z}) \big) :=
        \mathbb{E}_{\mathbf{z}_{i} \sim p(\mathbf{z})} \big[ \log{D(\mathbf{z}_{i})} \big] \\
        + \;  \mathbb{E}_{\mathbf{x}_{i} \sim p_{data}(\mathbf{x})} \big[ \log \big( 1 - D(E_{\theta}(\mathbf{x}_{i}) \big) \big]
    \end{split}
\end{equation}




\section{Molecular Hamiltonians}
\label{appendix:molecular_hamiltonians}

\subsection{Hydrogen molecule: $\text{H}_{2}$}
\label{sec:H2_molecule_appendix}

The two hydrogen atoms in the H$_{2}$ molecule are separated by interatomic distance $r$ and each atom is associated with a $1s$ atomic orbital. Therefore, the H$_{2}$ molecule has two molecular orbitals. 
The two molecular orbitals correspond to four spin orbitals and hence can be described by 4 qubits. We use the Jordan-Wigner fermion-to-qubit mapping to arrive at the general form of the qubit hamiltonian (Equation \ref{eq:h2_hamiltonian}) corresponding to the H$_{2}$ molecule \cite{qiskit_mappers_H2}.

\begin{multline}
     H(r) = c_0I + c_1(Z_0 +Z_2) + c_2(Z_1 + Z_3) + c_3(Z_1Z_0 + Z_3Z_2) \\
     + c_4(Z_2Z_0)
     + c_5(Z_3Z_0 +Z_1Z_2) + c_6Z_3Z_1 \\ + c_7(Y_3Y_2Y_1Y_0 + X_3X_2Y_1Y_0 + Y_3Y_2X_1X_0 + X_3X_2X_1X_0)
     \label{eq:h2_hamiltonian}
\end{multline}

The coefficients $\{ c_i \}$ in Equation \ref{eq:h2_hamiltonian} are a function of the interatomic distance $r$. The constant nuclear repulsion energy term 
is contained in the constant factor $c_{0}$.
However, using alternative mapping techniques 
that exploit the symmetries in the structure of $H(r)$, it is possible for $H(r)$ to have a simple 1-qubit representation as well: $H(r) = d_0I_0 + d_1Z_0 + d_2X_0$ \cite{qiskit_mappers_H2}.
Therefore, $H_{2}$ system can be simulated by using just 1-qubit for finding the ground state energy profile (blue curve in Figure \ref{H2_energy}). \\

Alternatively, one can also observe that the density matrices of the 4-qubit ground states of H$_{2}$ have a spare 1-qubit representation (an example depicted in Figure \ref{fig:h2_hm_1.5A}) and can therefore be compressed using a QAE. 
The sparsity in the ground state density matrix is due to the symmetries in the H$_{2}$ Hamiltonian (Equation \ref{eq:h2_hamiltonian}), allowing for a compression to a single qubit.

\subsection{Lithium Hydride: LiH}
\label{sec:LiH_molecule_appendix}

In the Lithium Hydride (LiH) molecule, the electron from the H atom occupies the $1s$ atomic orbital, while the LiH atom contributes three electrons: one in the $n=1$ shell (1 orbital) and two in the $n=2$ shell (4 orbitals). This results in a total of $1 + (1 + 4) = 6$ molecular orbitals, which correspond to 12 spin orbitals. Consequently, when applying a fermion-to-qubit mapping such as the Jordan-Wigner transformation, simulating the LiH molecule requires 12 qubits. \\

However, the size of this large system can be reduced to $6$ qubits by the employing following strategies:
\begin{enumerate}
    \item The core electrons in the $1s$ shell of Li do not contribute significantly to the chemistry of the LiH molecule. Therefore only the contribution from the valence shell of Li can be considered. This eliminates $2$ qubits corresponding to the core $1s$ orbital.

    \item The LiH molecule has rotational symmetry about one of the axes (say the $z$-axis). Therefore, along with freezing the inner $1s$ orbital, the contributions from its $p_{x}$ and $p_{y}$ orbitals can also be eliminated. This results in the further reduction of $(1 + 1) = 2$ molecular orbitals or $4$ spin orbitals (or qubits).

\end{enumerate}

Further, the Parity fermion-to-qubit mapping can be applied with two-qubit reduction to bring down the qubit count to $4$ qubits with certain loss in accuracy in the energy estimated. Different sequence of approximations and fermion-to-qubit mapping based on symmetries can also be applied to arrive at the $4$ qubit reduction \cite{qiskit_challenge_LiH}. Further elimination of qubits is not possible and this can be verified computationally \cite{qiskit_challenge_LiH}.


\section{Ansatzes used for the example of $\text{H}_{2}$}
\label{appendix:ansatz_H2}

\begin{figure*}[!thb]
    \centering
    \scalebox{0.8}{
        \begin{quantikz}
        \lstick[1]{$q_{0}$}&&
        \gategroup[5, steps=10,style={dashed, inner sep=6pt, rounded corners, xshift=+0.3cm}]{$\times \, d \, layers$}
        &  \gate{R_X(\theta_0)} & \gate{R_Y(\theta_n)} & & & \gate[nwires=1,style={fill=white,draw=white,text height=1cm}] {\ldots} & & \targ{} & \gate{R_X(\theta_{2n})} & \gate{R_Y(\theta_{3n})} & & \\
        \lstick[1]{\textit{$q_{1}$}}&  & & \gate{R_X(\theta_{1})} &\gate{R_Y(\theta_{n+1})} & & & \gate[nwires=1,style={fill=white,draw=white,text height=1cm}] {\ldots} & \targ{} & \ctrl{-1} & \gate{R_X(\theta_{2n+1})} & \gate{R_Y(\theta_{3n+1})} &&  \\
        \lstick[1]{\textit{\vdots}}&& & \gate[nwires=1,style={fill=white,draw=white,text height=1cm}]{\vdots} & \gate[nwires=1,style={fill=white,draw=white,text height=1cm}]{\vdots} & & \targ{} & \gate[nwires=1,style={fill=white,draw=white,text height=1cm}]{\vdots} & \ctrl{-1} & & \gate[nwires=1,style={fill=white,draw=white,text height=1cm}]{\vdots} & \gate[nwires=1,style={fill=white,draw=white,text height=1cm}]{\vdots} &&  \\
        \lstick[1]{\textit{$q_{n-2}$}}&& & \gate{R_X(\theta_{n-2})} &\gate{R_Y(\theta_{2n-2})} & \targ{} & \ctrl{-1} & \gate[nwires=1,style={fill=white,draw=white,text height=1cm}] {\ldots} & & & \gate{R_X(\theta_{3n-2})} & \gate{R_Y(\theta_{4n-2})} &&  \\
        \lstick[1]{\textit{$q_{n-1}$}}&& & \gate{R_X(\theta_{n-1})} &\gate{R_Y(\theta_{2n-1})} & \ctrl{-1} & & \gate[nwires=1,style={fill=white,draw=white,text height=1cm}] {\ldots} & & & \gate{R_X(\theta_{3n-1})} & \gate{R_Y(\theta_{4n-1})} &&  
    \end{quantikz}
}
    \caption{Structure of the ansatz used to implement the encoder/decoder to compress/reconstruct molecular ground states of $\text{H}_{2}$. Qubits $0$ to $l-1$ of such an ansatz for both the encoder and the decoder refer to the $l$-qubits of the latent state. The remaining $l$ to $n-1$ qubits refer to the \textit{trash}/\textit{auxiliary} qubits corresponding to the encoder/decoder. Each layer of this $n$-qubit ansatz of this structure has $4n$ tunable parameters for the rotation gates and $n-1$ CX gates. For the case of $\text{H}_{2}$, $l=1$, $n=4$ and $d=1$.}
    
    
    \label{fig:H2_qae_ansatz}
    
\end{figure*}
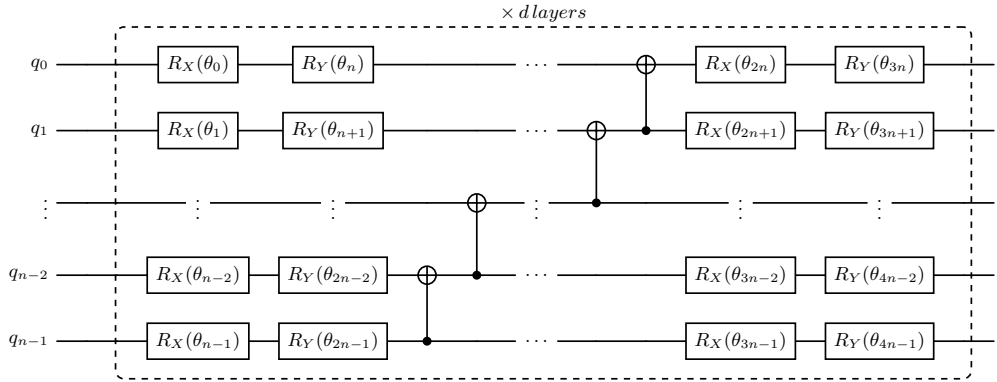

In this appendix, we present the structure of parametrized quantum circuits employed for the compression and reconstruction of the molecular ground states of $\mathrm{H}_{2}$. The choice of ansatz plays a critical role in determining the accuracy and efficiency of variational quantum algorithms. The specific ansatz circuits used in our simulations are illustrated in Figure~\ref{fig:H2_qae_ansatz}.


\section{Ansatzes used for the example of $\text{LiH}$}
\label{appendix:ansatz_LiH}

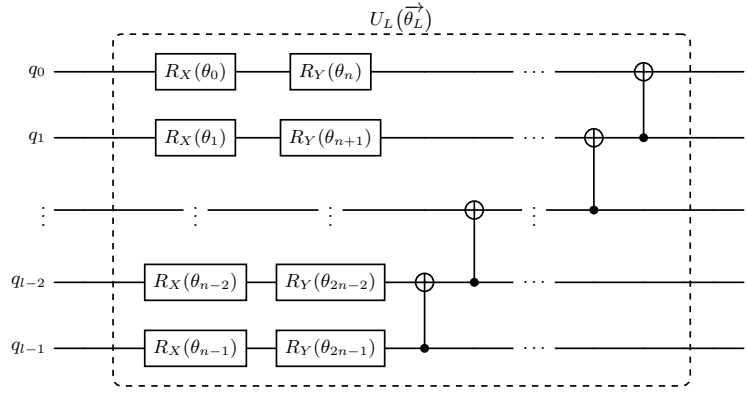
\begin{figure*}[!thb]
    \centering
    \scalebox{0.8}{
        \begin{quantikz}
        \lstick[1]{$q_{0}$}&&
        \gategroup[5, steps=8,style={dashed, inner sep=6pt, rounded corners, xshift=+0.3cm}]{$U_{L} \big (\mathbf{\overrightarrow{\theta_{L}}} \big)$}
        &  \gate{R_X(\theta_0)} & \gate{R_Y(\theta_n)} & & & \gate[nwires=1,style={fill=white,draw=white,text height=1cm}] {\ldots} & & \targ{} & & & \\
        \lstick[1]{\textit{$q_{1}$}}&  & & \gate{R_X(\theta_{1})} &\gate{R_Y(\theta_{n+1})} & & & \gate[nwires=1,style={fill=white,draw=white,text height=1cm}] {\ldots} & \targ{} & \ctrl{-1} &  &&  \\
        \lstick[1]{\textit{\vdots}}&& & \gate[nwires=1,style={fill=white,draw=white,text height=1cm}]{\vdots} & \gate[nwires=1,style={fill=white,draw=white,text height=1cm}]{\vdots} & & \targ{} & \gate[nwires=1,style={fill=white,draw=white,text height=1cm}]{\vdots} & \ctrl{-1} & &  &&  \\
        \lstick[1]{\textit{$q_{l-2}$}}&& & \gate{R_X(\theta_{n-2})} &\gate{R_Y(\theta_{2n-2})} & \targ{} & \ctrl{-1} & \gate[nwires=1,style={fill=white,draw=white,text height=1cm}] {\ldots} & & &  &&  \\
        \lstick[1]{\textit{$q_{l-1}$}}&& & \gate{R_X(\theta_{n-1})} &\gate{R_Y(\theta_{2n-1})} & \ctrl{-1} & & \gate[nwires=1,style={fill=white,draw=white,text height=1cm}] {\ldots} & & &  &&  
    \end{quantikz}
}
    \caption{The ansatz structure used to implement $U_{L} \big (\mathbf{\overrightarrow{\theta_{L}}} \big)$ in Figure \ref{fig:mol_gen_ansatz_LiH}. Each block of $U_{L} \big (\mathbf{\overrightarrow{\theta_{L}}} \big)$ has $2l$ tunable parameters for the rotation gates and $l-1$ CX gates. For the case of $\text{LiH}$, $l=4$.}
    
    
    \label{fig:LiH_gen_ansatz}
    
\end{figure*}

In this appendix, we present the quantum circuit ansatzes employed for the simulation of the Lithium Hydride (LiH) molecule. Figure~\ref{fig:LiH_mol_qae_ansatz} shows the structure of the ansatz used to implement the encoder and decoder for compressing and reconstructing the molecular ground states of LiH. In addition, Figure~\ref{fig:LiH_gen_ansatz} illustrates the ansatz structure used to implement $U_{L}(\bm{\theta_{L}})$ in Figure~\ref{fig:mol_gen_ansatz_LiH}. 

\begin{figure*}[!thb]
    \centering
    \scalebox{0.8}{
        \begin{quantikz}
        \lstick[1]{$q_{0}$}&&
        \gategroup[5, steps=9,style={dashed, inner sep=6pt, rounded corners, xshift=+0.3cm}]{$\times \, d \, layers$}
        &  \gate{R_X(\theta_0)} & \gate{R_Y(\theta_n)} & \ctrl{1} & & \gate[nwires=1,style={fill=white,draw=white,text height=1cm}] {\ldots} & &  & \targ{} &&\gate{R_X(\theta_{(2n)d})} & \gate{R_Y(\theta_{(3n)d})} & & \\
        \lstick[1]{\textit{$q_{1}$}}&  & & \gate{R_X(\theta_{1})} &\gate{R_Y(\theta_{n+1})} & \targ{} & \ctrl{1}& \gate[nwires=1,style={fill=white,draw=white,text height=1cm}] {\ldots} &  &  & &&\gate{R_X(\theta_{(2n+1)d})} & \gate{R_Y(\theta_{(3n+1)d})} &&  \\
        \lstick[1]{\textit{\vdots}}& & & \gate[nwires=1,style={fill=white,draw=white,text height=1cm}]{\vdots} & \gate[nwires=1,style={fill=white,draw=white,text height=1cm}]{\vdots} & & \targ{} & \gate[nwires=1,style={fill=white,draw=white,text height=1cm}]{\vdots} & \ctrl{1} & & &&\gate[nwires=1,style={fill=white,draw=white,text height=1cm}]{\vdots} &\gate[nwires=1,style={fill=white,draw=white,text height=1cm}]{\vdots} &&  \\
        \lstick[1]{\textit{$q_{n-2}$}}&& & \gate{R_X(\theta_{n-2})} &\gate{R_Y(\theta_{2n-2})} &  & & \gate[nwires=1,style={fill=white,draw=white,text height=1cm}] {\ldots} &\targ{}& \ctrl{1}& &&\gate{R_X(\theta_{(3n-2)d})} & \gate{R_Y(\theta_{(4n-2)d})} &&  \\
        \lstick[1]{\textit{$q_{n-1}$}}&& & \gate{R_X(\theta_{n-1})} &\gate{R_Y(\theta_{2n-1})} & & & \gate[nwires=1,style={fill=white,draw=white,text height=1cm}] {\ldots} & & \targ{}& \ctrl{-4}&&\gate{R_X(\theta_{(3n-1)d})} & \gate{R_Y(\theta_{(4n-1)})} &&  
    \end{quantikz}
}
    
    
    \caption{Structure of the ansatz used to implement the encoder/decoder to compress/reconstruct molecular ground states of $\text{LiH}$. Qubits $0$ to $l-1$ of such an ansatz for both the encoder and the decoder refer to the $l$-qubits of the latent state. The remaining $l$ to $n-1$ qubits refer to the \textit{trash}/\textit{auxiliary} qubits corresponding to the encoder/decoder. Each layer of this $n$-qubit ansatz of this structure has $2n$ tunable parameters for the rotation gates and $n$ circular entangling CX gates. For the encoder/decoder for the case of $\text{LiH}$, $l=4$, $n=6$ and $d=3$. The same ansatz structure is also used to implement the discriminator $U_{D}(K, \overrightarrow{\theta_{D}})$ in Figure \ref{fig:disc_structure} with $n = l + 1$ and $d = 3$.}
    
    
    
    \label{fig:LiH_mol_qae_ansatz}
    
\end{figure*}

\section{Caveats in the Quantum Autoencoder formalism} \label{QAE_caveats}

\begin{table*}[!tbh]
\centering
\newcolumntype{Y}{>{\centering\arraybackslash}X}

\renewcommand{\arraystretch}{1.4} 
\begin{tabularx}{\textwidth}{>{\centering\arraybackslash}p{1.8cm}|Y|>{\centering\arraybackslash}p{3cm}}
\hline
\textbf{} & \textbf{References} & \textbf{Compression Rate} \\ \hline
\multirow{5}{*}{\rotatebox[origin=c]{90}{\textbf{HARDWARE}}} 
& Realization of a quantum autoencoder for lossless compression of quantum data. \cite{huang2020realization} & 2 $\rightarrow$ 1 \\
& Experimental Realization of a Quantum Autoencoder: The Compression of Qutrits via Machine Learning. \cite{pepper2019experimental} & Qutrits $\rightarrow$ Qubits \\
& Implementation of quantum compression on IBM quantum computers. \cite{pivoluska2022implementation} & 3 $\rightarrow$ 2 \\
& Information loss and run time from practical application of quantum data compression. \cite{Patel_2023} & 4 $\rightarrow$ 1 \\
& Quantum autoencoders using mixed reference states. \cite{ma2024quantum} & 4 $\rightarrow$ 2 \\ \hline

\multirow{6}{*}{\rotatebox[origin=c]{90}{\textbf{SIMULATION}}} 
& Variational Denoising for Variational Quantum Eigensolver. \cite{PhysRevResearch.6.023181} & Up to 14 qubits \\
& Quantum autoencoders for efficient compression of quantum data. \cite{romero2017quantum} & Up to 8 qubits \\
& &  \\
& &  \\
& &  \\
& &  \\
\hline
\end{tabularx}

\caption{Summary of Quantum Autoencoder Studies and Their Compression Capabilities}
\label{table:qubits}
\end{table*}

The compression capabilities of QAEs are subject to fundamental constraints such as structural properties, entanglement, and entropy of the input states. Theoretical investigations have demonstrated that QAEs can achieve lossless compression of high-dimensional quantum states into a lower-dimensional latent space~\cite{huang2020realization, qae_compression_bounds}. A necessary condition for such lossless compression is that the number of linearly independent input states does not exceed the dimensionality of the target latent space~\cite{qae_compression_bounds}.

Despite their potential, QAEs face several challenges in scalability and implementation. Current experimental demonstrations are restricted to small systems due to hardware limitations such as noise, shallow circuit depths, and limited qubit connectivity, all of which reduce reconstruction fidelity~\cite{pepper2019experimental, pivoluska2022implementation, Patel_2023, qae_mixed_reference_states}. QAE performance is sensitive to design choices including circuit depth, compression ratio, and the choice of parameterized unitaries~\cite{Patel_2023}. Additionally, large-scale deployment of QAEs may require deep quantum circuits and exponential classical resources for tasks like fidelity or entropy estimation (e.g., via SWAP tests~\cite{swap_quantum_fingerprinting} or full state tomography ~\cite{cramer2010efficient}), resulting in significant overhead that exceeds current NISQ-era capabilities. Classical shadow methods offer a potential means to partially alleviate these challenges by enabling more efficient estimations; however, their effectiveness in this context has yet to be thoroughly tested \cite{huang2020predicting}. Table~\ref{table:qubits} summarizes existing implementations on quantum hardware and simulators, along with their corresponding compression rates.

\section{Additional QAE experiments on LiH}

\label{LiH_QAE_add}

We additionally consider a necessary condition for perfect quantum state compression: the number of linearly independent input states must not exceed the dimensionality of the target latent space \cite{huang2020realization,qae_compression_bounds}. In the case of pure quantum states, each individual state is rank-one. However, when dealing with a dataset comprising multiple pure states, the effective rank is determined by the span of these states, which corresponds to the rank of the density matrix formed by the ensemble of pure states. In contrast, if the dataset contains mixed states with some prior distribution, the relevant rank is that of the averaged density matrix describing the ensemble. This rank captures both the intrinsic mixedness of each state and the statistical mixture induced by the prior, and it sets the minimum latent dimension required for lossless compression.


\begin{figure*}[!bht]
    \centering
    \begin{subfigure}{0.49\textwidth}
       \includegraphics[width=\textwidth]{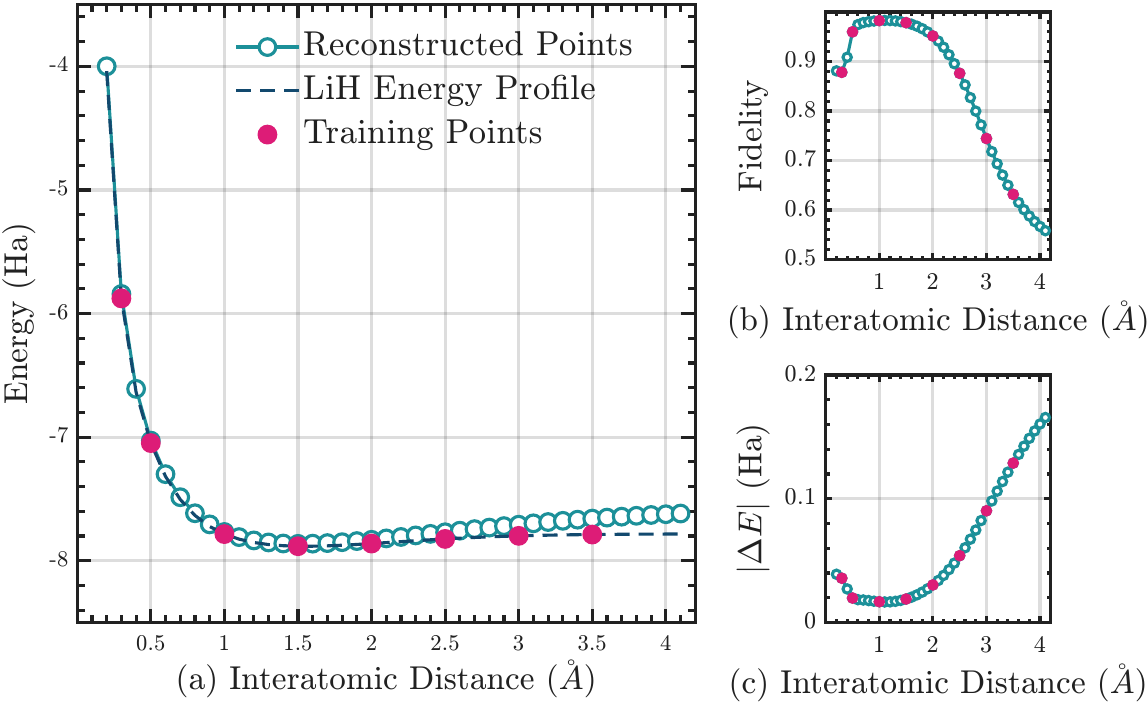}
        \caption{LiH (10 qubits)}
        \label{fig:LiH_QAE_10}
    \end{subfigure}
    \begin{subfigure}{0.49\textwidth}
        \includegraphics[width=\textwidth]{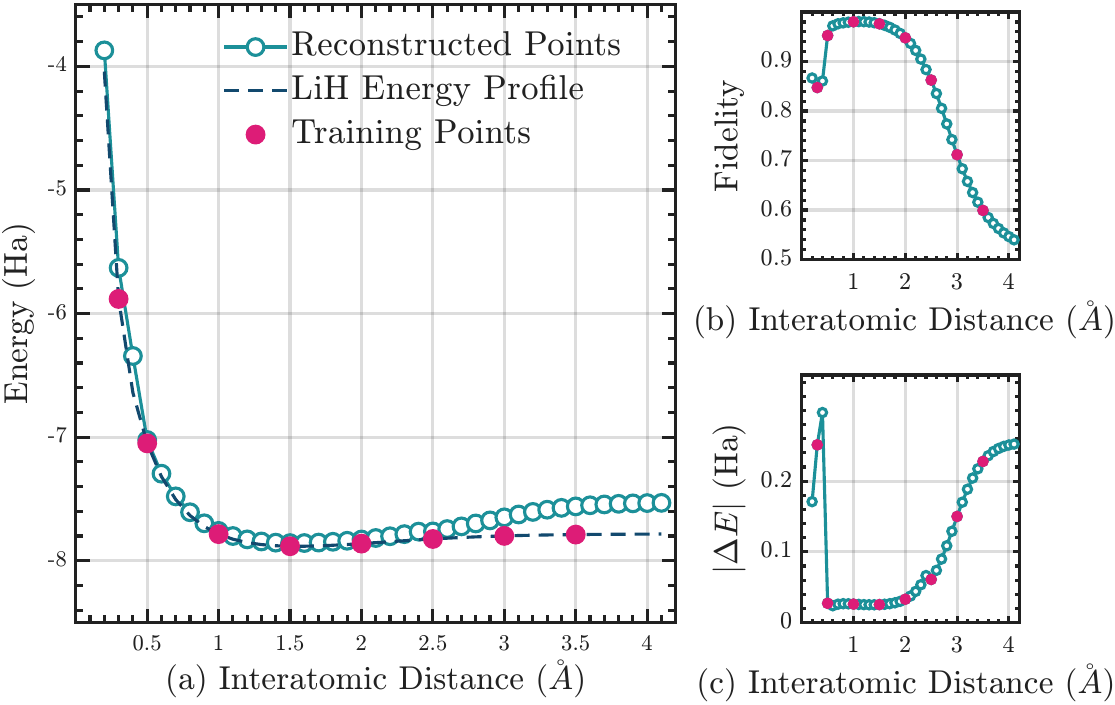}
        \caption{LiH (12 qubits)}
        \label{fig:LiH_QAE_12}
    \end{subfigure}
    \caption{Ground state energy profile across different interatomic distances of the molecules. The red points denote the ground states that were selected as training data for the QAE. The key takeaway is that, by learning from these representative configurations, the QAE can ideally reduce the ground states of LiH molecules from 12 or 10 qubits to a significantly smaller latent space while retaining essential physical information.}
    \label{fig:LiH_QAE_10_12}
\end{figure*}

Considering the LiH molecule, where the dataset is generated by varying the interatomic distance from $0.2$ to $4.2$ Hartree, in steps of $0.1$ Hartree. Each configuration yields a ground state, represented as a pure state. Assuming an equal probability distribution over all configurations, the ensemble can be used to construct a density matrix. The rank of this density matrix reflects the number of linearly independent pure states in the dataset. We compute the rank of the input state ensemble for Hamiltonians of $12$, $10$, and $6$ qubits, obtaining ranks of $18$, $7$, and $6$, respectively. These results suggest that the latent space can be represented using $5$ qubits for the $12$-qubit Hamiltonian, and $3$ qubits for both the $10$- and $6$-qubit Hamiltonians for lossless compression. While this approach provides a guideline for selecting the number of qubits in the latent space, the exact computation of the rank scales exponentially with the size of the input Hilbert space. This renders the method computationally intensive and challenging to scale for larger systems.

\textbf{Compressing LiH ground State:}
We employ QAE to compress the ground states of the LiH molecule for systems initialized with 12 and 10 qubits. Table \ref{table:qae} and Figure \ref{fig:LiH_QAE_10_12} present the reconstructed energy profiles along with the corresponding fidelity and reconstruction error for both cases.

\begin{table}[!tbh]
    \centering

    \begin{tabular}{c|c|c|c}

        \toprule
 \textbf{Molecule} & Compression Rate & $\big \langle \mathcal{F}(\sigma_{r}, \rho_{r}) \big \rangle$ & {$\big \langle |\Delta E (r)| \big \rangle$ } \\
        \midrule

         $\text{LiH}$ &  $10$ $\longrightarrow$ $4$ qubits &   $ 0.842
 \pm 0.148$ & $0.062 \pm 0.049$ Ha \\

         $\text{LiH}$ & $12$ $\longrightarrow$ $4$ qubits &  $ 0.826
 \pm 0.159$ & $0.111 \pm 0.093$ Ha \\

        \bottomrule

    \end{tabular}

\caption{Performance Metrics of QAE for compression and reconstruction of Molecular Ground States.}
\label{table:qae}
\end{table}

\section{QGAA vs QGAN: Training the QGAN directly on the 4 qubit ground states of H$_2$} 
\label{QGANvsQGAA}
\begin{figure}[!tbh]
    \includegraphics[width=0.5\textwidth]{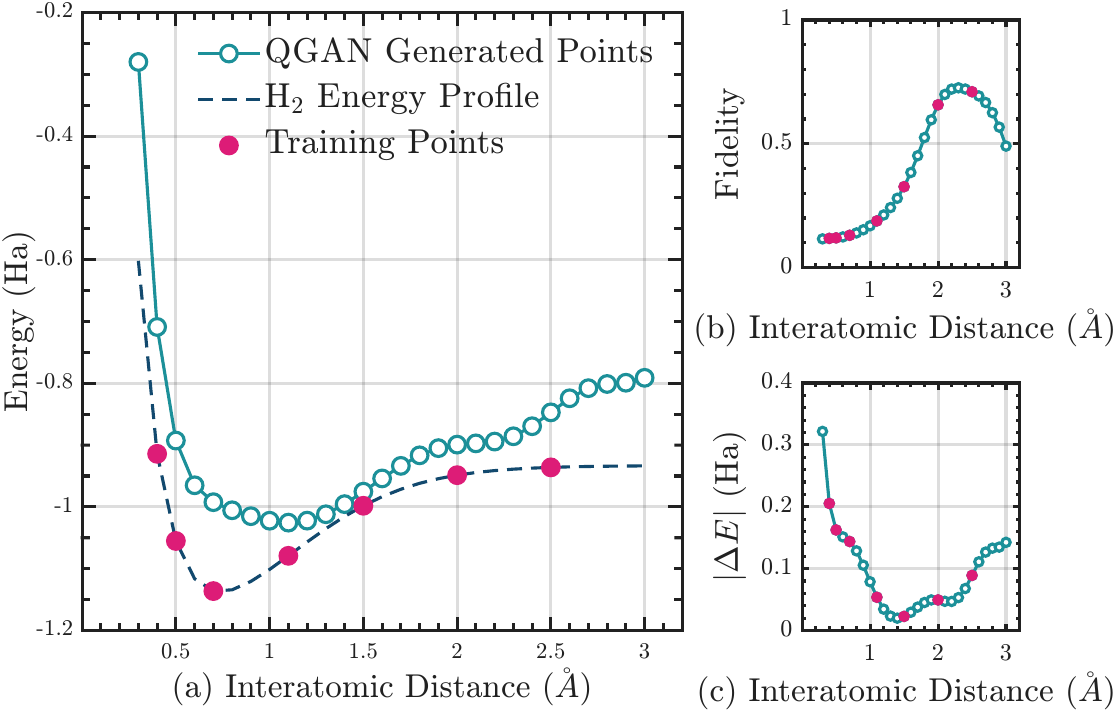}
    
    \label{fig:H_2_qgan_4}

    \caption{Learned energy profiles after adversarial training on 4 qubits input space (without compression using QAE). Each plot also includes the fidelity and reconstruction error of the generated states as a function of interatomic distance. The QGAN learns the energy profile whereas the QGAA achieves better precision.}
    \label{fig:qgan_energy_profiles_appen}
\end{figure}

In this section, we compare the performance of the Quantum Generative Adversarial Network (QGAN) when trained directly on the $4$-qubit ground states of the hydrogen molecule (H$_2$) with the Quantum Generative Autoencoder Architecture (QGAA), which is trained on a compressed latent representation of these ground states. While the QGAA leverages an autoencoding step to reduce the dimensionality of quantum states before adversarial training, the QGAN attempts to learn the full ground state distribution without compression. This comparison highlights the trade-offs between direct generative modeling of high-dimensional quantum states and the efficiency gains achieved by incorporating an autoencoder into the training pipeline. As shown in Fig.~\ref{fig:qgan_energy_profiles_appen}, we present the learned energy profile obtained by training the QGAN on $4$-qubit ground states, i.e., without employing the autoencoding step used in the QGAA.


\end{document}